\documentclass[12pt]{article}
\pdfoutput=1

\usepackage{amsfonts,amssymb,cancel,hyperref,mathtools,multirow,setspace}
\usepackage[textheight=9in,textwidth=7in,letterpaper]{geometry}

\numberwithin{equation}{section}
 
\newcommand{\ab}[1]{\left|#1\right|}

\newcommand{\br}[1]{\left[#1\right]}
\newcommand{\cu}[1]{\left\{#1\right\}}
\newcommand{\pa}[1]{\left(#1\right)}

\newcommand{\ed}{\,\mathrm{d}}
\newcommand{\pd}{\,\partial}

\newcommand{\J}{\mathcal{J}}
\renewcommand{\L}{\mathcal{L}}
\renewcommand{\P}{\mathcal{P}}

\newcommand{\N}{\mathbb{N}}
\newcommand{\Z}{\mathbb{Z}}
\newcommand{\R}{\mathbb{R}}
\newcommand{\C}{\mathbb{C}}

\newcommand{\SO}{\mathsf{SO}}
\newcommand{\SL}{\mathsf{SL}}
\newcommand{\U}{\mathsf{U}}

\begin{document}

\hfill

\begin{center}
	\vspace{1cm}
	{\LARGE\bf Exact Solutions for}\\
	\vspace{2mm}
	{\LARGE\bf Extreme Black Hole Magnetospheres}\\
	\vspace{2cm}
	Alexandru Lupsasca$^\spadesuit$ and Maria J. Rodriguez$^\heartsuit$\\
	\vspace{1cm}
	{\it Center for the Fundamental Laws of Nature, Harvard University\\
	Cambridge, MA 02138, USA}
\end{center}

\vspace{2cm}
\footnotetext{$^\spadesuit$lupsasca@fas.harvard.edu
$\quad^\heartsuit$mjrodri@physics.harvard.edu}

\begin{abstract}
	We present new exact solutions of Force-Free Electrodynamics (FFE) in the Near-Horizon region of an Extremal Kerr black hole (NHEK) and offer a complete classification of the subset that form highest-weight representations of the spacetime's $\SL(2,\R)\times\U(1)$ isometry group. For a natural choice of spacetime embedding of this isometry group, the $\SL(2,\R)$ highest-weight conditions lead to stationary solutions with non-trivial angular dependence, as well as axisymmetry when the $\U(1)$-charge vanishes. In addition, we unveil a hidden $\SL(2,\C)$ symmetry of the equations of FFE that stems from the action of a complex automorphism group, and enables us to generate an $\SL(2,\C)$ family of (generically time-dependent) solutions. We then obtain still more general solutions with less symmetry by appealing to a principle of linear superposition that holds for solutions with collinear currents. This allows us to resum the highest-weight primaries and their $\SL(2,\R)$-descendants.
\end{abstract}

\vspace{3cm}
\pagebreak

\tableofcontents

\vspace{.5cm}

\section{Introduction and Summary}
\label{sec:Introduction}

The electromagnetic field inside a magnetically-dominated relativistic plasma is generically described by the nonlinear equations of Force-Free Electrodynamics (FFE). These equations are believed to capture the behavior of a wide range of astrophysical systems, ranging from neutron stars to active black holes \cite{Michel:1973,Blandford:1977ds}. Force-free magnetospheres typically surround these objects, powering some of the most extravagant energy signals in the universe: pulsars (in the case of neutron stars) and active galactic nuclei, or quasars (in the case of black holes). FFE is expected to play a crucial role in efforts to elucidate the origin of such signals.

Though the significance of FFE to our understanding of magnetospheres was recognized several decades ago, it is only recently that the subject has attracted broader interest beyond the astrophysics community. The equations of FFE naturally arise in a variety of settings, from the modeling of the magnetic fields of our own Sun \cite{Wiegelmann:2012mu} to the realization last year of a force-free plasma in a laboratory on Earth \cite{PhysRevLett.110.085002}. Because the equations of FFE characterize ubiquitous physical phenomena, we hold that they should join the list of nonlinear equations considered fundamental in physics, a list that already includes the Einstein equations of gravity and the Navier-Stokes equations of fluid dynamics. In this spirit, the theory of FFE certainly merits investigation in its own right.

Few analytic solutions of FFE have been found \cite{Michel:1973,Blandford:1976,Blandford:1977ds,Menon:2007,Tanabe:2008wm,Brennan:2013jla,Brennan:2013ppa,Gralla:2014yja,Zhang:2014pla}. Recently \cite{Lupsasca:2014pfa}, a symmetry-based approach to FFE led to the discovery of several infinite families of force-free solutions in the Near-Horizon Region of Extreme Kerr black holes (NHEK). This geometry, which forms a spacetime in its own right (albeit a non-asymptotically flat one), presents an excellent opportunity to develop such an approach because it enjoys an enlarged isometry group: in NHEK, the $\U(1)$ time symmetry of the Kerr black hole is enhanced to an $\SL(2,\R)$ conformal group \cite{Bardeen:1999px}, a symmetry group whose exploitation has already proved fruitful in the context of the Kerr/CFT correspondence \cite{Guica:2008mu,Bredberg:2011hp}.

In this paper, we pursue the line of investigation opened in \cite{Lupsasca:2014pfa} and apply the symmetry-based approach to FFE in the NHEK region more systematically. Along the way, we begin to uncover the beautiful structure of the force-free equations and develop three new solution-generating techniques: 
\begin{itemize}
\item
First, one may apply a nonlinear transformation to known force-free solutions to obtain new solutions with current flowing in new directions. This mechanism, which we describe in section \ref{sec:TypeEM}, relies on the nonlinear superposition of the purely magnetic and purely electric solutions derived in sections \ref{sec:TypeM} and \ref{sec:TypeE}, and yields electromagnetic solutions with current flowing in the $\theta$-direction, $\J_\theta\neq0$.
\item
Second, one may consider the transformation properties of the solutions under $\SL(2,\C)$, the complexification of the $\SL(2,\R)$ isometry group. While the action of $\SL(2,\R)$ does not provide new solutions (but rather produces solutions related by finite $\SL(2,\R)$ transformations), the complex $\SL(2,\C)$ transformations yield new solutions that are related to the original ones by complex diffeomorphisms, and are therefore physically inequivalent. We work out many of the details in section \ref{sec:Automorphisms}.
\item
Third, one may exploit the previously-observed \cite{Lupsasca:2014pfa} phenomenon of linear superposition within certain infinite families of solutions to obtain more general solutions with less symmetry. We explain the origin of this phenomenon in appendix \ref{appendix:Collinearity} and apply it to $\SL(2,\R)$-descendants in section \ref{sec:Resummation}. The $\SL(2,\R)$ highest-weight solutions and their descendants may be construed as a mode expansion in conformal harmonics, akin to a spherical expansion. This indicates the existence of solutions with full functional freedom, which we finally present in section \ref{sec:RealSolutions}.
\end{itemize}
These techniques enable us to derive the large class of solutions presented in Table \ref{table:RealSolutions}, including some that describe magnetically-dominated ($F^2\propto\vec{B}^2-\vec{E}^2>0$) stationary axisymmetric magnetospheres.

The outline of the paper is as follows. In section \ref{sec:NHEK}, we briefly review the geometry and symmetries of the NHEK region. Readers who are familiar with these details and are mostly interested in explicit solutions may skip this section. One key difference with \cite{Lupsasca:2014pfa} is a change of the spacetime embedding of the isometry group. We now adopt a more natural choice from the perspective of the Poincar\'e observer (which is the one descended from the Boyer-Lindquist observer in the scaling limit), and obtain precisely three families of solutions: a purely Magnetic Type, a purely Electric Type, and a mixed Electromagnetic Type. These are quickly derived from the ground up in Sections \ref{sec:TypeM}, \ref{sec:TypeE} and \ref{sec:TypeEM}, respectively. They all exhibit a complicated $\theta$-dependence that we analyze in appendix \ref{appendix:ThetaAnalysis}.

Subsequently, in section \ref{sec:Automorphisms} we revisit the choice of embedding, and determine that it is paramaterized by a complexified $\SL(2,\C)$ isometry group. One of its $\SL(2,\R)$ subgroups is merely part of the isometry group and therefore acts trivially on the solutions, but the other ``hidden" component of $\SL(2,\C)$ generates highly nontrivial time-dependent solutions.

\clearpage

\begin{table}[t!]
\centering
\resizebox{\columnwidth}{!}{
\begin{tabular}{| c | c | c | c |}
\hline
	\multicolumn{1}{|c|}{Vector potential $A$} & \multicolumn{3}{c|}{Properties}  \\
\hline
	& & & \\
	$\displaystyle\frac{1}{2J\Gamma}\br{\frac{\Phi}{\sqrt{2}}H_+-\pa{1-\frac{1}{\Lambda^2}}W_0}\int\!\ed h\,c_h\Phi^hS_h$
	& \multirow{18}{*}{\rotatebox{-90}{\!\!Stationary~~~~}}
	& \multirow{9}{*}{\rotatebox{-90}{\!\!Axisymmetric~~}}
	& \multirow{26}{*}{\rotatebox{-90}{$F^2\neq0$~~}} \\
	& & & \\
\cline{1-1}
	& & & \\
	$\displaystyle\frac{1}{2J\Gamma}\br{\frac{\Phi}{\sqrt{2}}H_+-W_0}\int\!\ed h\,c_h\Phi^hP_h$
	& & & \\
	& & & \\
\cline{1-1}
	& & & \\
	$\displaystyle\frac{\Phi^hU_h}{2J\Gamma}\br{
		\frac{\Phi}{\sqrt{2}}\pa{1+\frac{DhU_h^{-1/h}}{h-1}}H_+
		-\pa{1-\frac{1}{\Lambda^2}+\frac{DhU_h^{-1/h}}{h-1}}W_0+\frac{CU_h^{-1/h}}{\Lambda}\Theta}$
	& & & \\
	& & & \\
\cline{1-1}\cline{3-3}
	& & & \\
	$\displaystyle\frac{1}{2J\Gamma}\int\!\ed h\ed m\,c_h\Phi^h
	\cu{b_mS_{h,m}\br{\frac{\Phi}{\sqrt{2}}H_+-\pa{1-\frac{1}{\Lambda^2}}W_0}
		-b_m'S_{h,m}'\chi_{h,m}\pa{1-\frac{1}{\Lambda^2}}\Theta}$
	& & \multirow{6}{*}{\rotatebox{-90}{\!\!\!\!\!\!\!\!\!\!\!\!Non-Axisymmetric}} & \\
	& & & \\
\cline{1-1}
	& & & \\
	$\displaystyle\frac{\Phi^h}{2J\Gamma}\pa{\frac{\Phi}{\sqrt{2}}H_+-W_0}\int\!\ed m\,b_mP_m\qquad\pa{h\in\cu{0,1}}$
	& & & \\
	& & & \\
\cline{1-3}
	& & & \\
	$\displaystyle\frac{\Phi^hP_h}{2J\Gamma}\br{\Phi\pa{\frac{h+1}{\sqrt{2}}tH_+-H_0}-htW_0}\int\!\ed h\br{C\delta(h)+D\delta(1-h)}$
	& \multirow{9}{*}{\rotatebox{-90}{Time-dependent~~}}
	& \multirow{9}{*}{\rotatebox{-90}{\!\!Axisymmetric~~}} & \\
	& & & \\
\cline{1-1}
	& & & \\
	$\displaystyle\frac{1}{2J\Gamma}g\pa{t\pm\frac{1}{r}}\br{\frac{C}{\Lambda}\Theta+\frac{D}{\Lambda^2}W_0}$
	& & & \\
	& & & \\
\cline{1-1}\cline{4-4}
	& & & \multirow{4}{*}{\rotatebox{-90}{\!\!$F^2=0$~~}} \\
	$\displaystyle\frac{1}{2J\Gamma}f\pa{t\pm\frac{1}{r},\theta}\Theta$
	& & & \\
	& & & \\
\hline
	& \multirow{4}{*}{\rotatebox{-90}{\!\!Modified~~}} & \multicolumn{2}{c|}{\multirow{4}{*}{\rotatebox{-90}{\!\!Same~~}}} \\
	$\SL(2,\C)$-transformed analogues of the above solutions & & \multicolumn{2}{c|}{} \\
	(all time-dependent) & & \multicolumn{2}{c|}{} \\
	& & \multicolumn{2}{c|}{} \\
\hline
\end{tabular}}
\caption{Vector potentials $A$ that define solutions $F=\ed A$ to the equations \eqref{eq:FFE1}--\eqref{eq:FFE3} of FFE in NHEK. These potentials are constructed from highest-weight solutions with respect to the Poincar\'{e} basis $\cu{H_0,H_\pm}$ of $\SL(2,\R)$ (defined in section \ref{sec:NHEK}) using various techniques developed in the paper. Here, $c_h\Phi^h(r)=c(h)r^{-h}$ and $b_m(\phi)=x(m)\cos{m\phi}+y(m)\sin{m\phi}$, with $c(h),x(m),y(m)$ arbitrary real functions, while $C$ and $D$ are arbitrary real coefficients. The $\theta$-dependent functions $P_h,P_m,S_h,S_{h,m},\chi_{h,m}$ are defined throughout the text. Analogous solutions can be derived in the other bases of $\SL(2,\R)$ using a hidden $\SL(2,\C)$ symmetry, as outlined in section \ref{subsec:Survey}.}
\label{table:RealSolutions}
\end{table}

\clearpage

After this generalization, the three aforementioned families no longer have a definite electric or magnetic type, and are more appropriately termed Type M, Type E and Type EM, respectively. As an illustrative example, we show that for certain values of the hidden $\SL(2,\C)$ parameters, our Type E solutions reduce to the family of non-null solutions presented in \cite{Lupsasca:2014pfa}. The full classification of highest-weight solutions, labeled by $\SL(2,\C)$ parameters, is presented in Table \ref{table:Classification}.

In section \ref{sec:Resummation}, we investigate $\SL(2,\R)$-descendants and determine the conditions under which they may admit a principle of linear superposition. Invoking the theorems derived in Appendix \ref{appendix:Collinearity}, we obtain a quick criterion for resummability, and then form the most general linear combinations of solutions. These are in general complex, but nonetheless their real and imaginary parts are often still solutions due to the collinearity of their currents. We can then list the most general physical solutions amenable to discovery through our symmetry-based approach, be they stationary or time-dependent, axisymmetric or not. Table \ref{table:RealSolutions} summarizes the outcome of this procedure.

The real force-free solutions we obtain contain at most two free functions, compared with the four independent functions required to specify initial data for the Cauchy problem in FFE. The initial data consists of six components of the electric and magnetic fields $\vec{E},\vec{B}$, minus two degrees of freedom removed by the constraints $\vec{\nabla}\cdot\vec{B}=0$ and $\vec{E}\cdot\vec{B}=0$ \cite{Komissarov:2002,Pfeiffer:2013wza}. Hence there likely exist other force-free solutions in NHEK beyond the ones we have found using our symmetry-based approach.

Interestingly, we find that several of these solutions contain free functions on an AdS$_2$ subfactor of NHEK (which may be viewed as warped fibration of WAdS$_3$ over $S^1$). Only one of them is null ($F^2=0$), so it should therefore reproduce the null solutions previously found in \cite{Lupsasca:2014pfa} -- we check this is indeed the case in section \ref{subsec:Survey}.

Finally, in section \ref{subsec:nearNHEK} we examine conformal transformations from NHEK to near-NHEK, the bulk geometry of a near-horizon near-extremal Kerr black hole. Such a black hole acquires a (low) temperature $\kappa>0$ due to its (small) deviation from extremality. All of our solutions remain exactly valid in near-NHEK. Moreover, we show that by a judicious choice of conformal transformation, the horizon singularities of our solutions in NHEK may be pushed beyond the near-NHEK horizon.

At present, it is still unclear how to extend force-free solutions in (near-)NHEK beyond the near-horizon region and into the full Kerr spacetime. This important question remains open.

\section{NHEK geometry}
\label{sec:NHEK}

In this section we briefly review the Kerr black hole, as well as the geometry and symmetries of the NHEK region. We also lay out the symmetry approach to the force-free equations in NHEK.

\subsection{Kerr black hole}

The Kerr metric models astrophysically realistic rotating black holes. In Boyer-Lindquist coordinates $(\hat t,\hat r,\hat\theta,\hat\phi)$ with natural units (we set $c=G=1$), its line element is
\begin{align}
	\label{eq:KerrMetric}
	ds^2=&-\frac{\Delta}{\Sigma}\pa{\ed\hat t-a\sin^2{\hat\theta}\ed\hat\phi}^2
		+\frac{\Sigma}{\Delta}\ed\hat r^2+\frac{\sin^2\hat\theta}{\Sigma}\br{\pa{\hat r^2+a^2}\ed\hat\phi
		-a\ed\hat t}^2+\Sigma\ed\hat\theta^2.
\end{align}
Here $J$ parametrizes the angular momentum of the black hole, $M$ its mass, and
\begin{align}
	\Delta=\hat r^2-2M\hat r+a^2,\qquad
	\Sigma=\hat r^2+a^2\cos^2\hat\theta,\qquad
	a=\frac{J}{M}.
\end{align}
There is an event horizon at
\begin{align}
	\hat r_H=M+\sqrt{M^2-a^2},
\end{align}
from which it follows that the Kerr solution has a naked singularity unless $\ab{a}\le M$. This last bound is saturated by the so-called extreme Kerr solution, which carries the maximal angular momentum
\begin{align}
	\ab{J}=M^2.
\end{align}
As long as $J>0$, the Kerr black hole is spinning and develops an ergosphere, which is bounded by the hypersurface
\begin{align}
	\hat r=M+\sqrt{M^2-a^2\cos^2{\hat\theta}}.
\end{align}

\subsection{NHEK region: the scaling limit}

In this paper we are interested in the region very close to the horizon of extreme Kerr. It is described by the so-called Near-Horizon Extreme Kerr (NHEK) geometry, which can be obtained by a near-horizon limiting procedure from the Kerr metric in usual Boyer-Lindquist coordinates \eqref{eq:KerrMetric}. Following \cite{Bardeen:1999px}, define new dimensionless coordinates $(t,r,\theta,\phi)$ by 
\begin{align}
	t=\frac{\lambda\hat t}{2M},\qquad
	r=\frac{\hat r-M}{\lambda M},\qquad
	\theta=\hat\theta,\qquad
	\phi=\hat\phi-\frac{\hat t}{2M}.
\end{align}
In taking the limit $\lambda\to0$ while keeping these coordinates fixed, one is effectively ``zooming" into the region near the horizon. This procedure yields the NHEK line element in Poincar\'{e} coordinates
\begin{align}
	ds^2=2J\Gamma\br{-r^2\ed t^2+\frac{\ed r^2}{r^2}+\ed\theta^2+\Lambda^2(\ed\phi+r\ed t)^2},
\end{align}
where $t\in(-\infty,\infty)$, $r\in[0,\infty)$, $\theta\in[0,\pi]$, $\phi\sim\phi+2\pi$ and
\begin{align}
	\Gamma(\theta)=\frac{1+\cos^2\theta}{2},\qquad
	\Lambda(\theta)=\frac{2\sin\theta}{1+\cos^2\theta}.
\end{align}
This metric may be viewed as a warped fibration (warped by $\Lambda$) of WAdS$_3$ over the $S^1$ parameterized by $\theta$, with the $(t,r)$ coordinates forming an AdS$_2$ subfactor. In contrast with the original Kerr metric \eqref{eq:KerrMetric}, the NHEK geometry is not asymptotically flat. The event horizon of the original extreme Kerr black hole is now located at
\begin{align}
	\label{eq:EventHorizon}
	r_H=0.
\end{align}
The boundary of the ergosphere reduces to two hypersurfaces of constant $\theta=\theta_e^\pm$, with
\begin{align}
	\label{eq:Ergosphere}
	\Lambda\pa{\theta_e^\pm}=1\qquad\Longleftrightarrow\qquad
	\theta_e^\pm=\frac{\pi}{2}\pm\br{\arcsin\pa{\sqrt{3}-1}-\frac{\pi}{2}}.
\end{align}
This boundary is an AdS$_3$ subfactor of NHEK.

\subsection{Isometries of NHEK}

Under the scaling limit described in the previous section, the original $\U(1)\times\U(1)$ Kerr isometry group is enlarged to an $\SL(2,\R)\times\U(1)$ symmetry that governs the dynamics of the NHEK region. The $\U(1)$ rotational symmetry is generated by the Killing vector field
\begin{align}
	W_0=\pd_\phi.
\end{align}
The time translation symmetry becomes part of an enhanced $\SL(2,\R)$ isometry group generated by the Killing vector fields
\begin{align}
	H_0&=t\pd_t-r\pd_r,\\
	H_+&=\sqrt{2}\pd_{t},\\
	H_-&=\sqrt{2}\br{\frac{1}{2}\pa{t^2+\frac{1}{r^2}}\pd_t-tr\pd_r-\frac{1}{r}\pd_{\phi}}.
\end{align}
It is easily verified that these  satisfy the $\SL(2,\R)\times\U(1)$ commutation relations, namely:
\begin{align}
	\br{H_0,H_\pm}&=\mp H_\pm,\qquad\,\br{H_+,H_-}=2H_0,\\
	\br{W_0,H_\pm}&=0,\qquad\qquad\br{W_0,H_0}=0.
\end{align}
These symmetries do not leave the original Kerr horizon \eqref{eq:EventHorizon} invariant and mix up the inside and outside of the original black hole. Note that this parameterization of $\SL(2,\R)$ is natural from the perspective of the Poincar\'e observer, and differs from the global basis $\cu{L_0,L_\pm}$ employed in \cite{Lupsasca:2014pfa} -- we will return to this crucial point in section \ref{sec:Automorphisms}. Here and hereafter, it is understood that in this paper we use the same symbol (e.g. `$H_+$') to denote both a vector field and its associated 1-form, and rely on the context to distinguish between the two uses.

\subsection{FFE in NHEK}

We wish to solve the force-free equations for a 2-form $F$ in the background of NHEK:
\begin{align}
	\label{eq:FFE1}
	\ed F&=0,\\
	\label{eq:FFE2}
	\ed^\dagger F&=\J,\\
	\label{eq:FFE3}
	\star F\wedge\J&=0.
\end{align}
To exploit the isometries of the NHEK region, we focus on solutions that lie in highest-weight representations of $\SL(2,\R)$ and carry $\U(1)$-charge. That is, we assume that
\begin{align}
	\label{eq:HW1}
	\L_{H_+}F&=0,\\
	\label{eq:HW2}
	\L_{H_0}F&=hF,\\
	\label{eq:HW3}
	\L_{W_0}F&=imF.
\end{align}
The most general 2-form satisfying these conditions is best expressed in tensor notation as
\begin{align}
	F_{\mu\nu}=\frac{e^{im\phi}}{r^h}
	\begin{pmatrix}
		0 & A(\theta) & rB(\theta) & rC(\theta) \\
		& 0 & X(\theta)/r & Y(\theta)/r \\
		& & 0 & Z(\theta) \\
		& & & 0
	\end{pmatrix}
	.
\end{align}
It is an angular momentum eigenstate with angular momentum $m$, and forms a highest-weight representation of $\SL(2,\R)$ with highest weight $h$. Evidently, we must require that $m\in\Z$ to ensure that $F$ is not multi-valued. This group-theoretic Ansatz for $F$ completely fixes its dependence on the 3 coordinates $(t,r,\phi)$, thereby leaving it undetermined only up to six arbitrary functions of $\theta$. Imposing the Bianchi identity \eqref{eq:FFE1},
\begin{align}
	\ed F=0,
\end{align}
eliminates half of these functions:
\begin{align}
	A(\theta)&=\frac{i}{m}(h-1)C(\theta),\\
	B(\theta)&=-\frac{i}{m}C'(\theta),\\
	X(\theta)&=-\frac{i}{m}\br{hZ(\theta)+Y'(\theta)}.
\end{align}
Thus, before even plugging this Ansatz into the force-free equations proper, $F$ is already fixed up to three scalar functions of $\theta$. Furthermore, recall that a 2-form $F$ that solves the force-free equations \eqref{eq:FFE1}--\eqref{eq:FFE3} must necessarily be degenerate \cite{Gralla:2014yja}, in the sense that
\begin{align}
	F\wedge F=0.
\end{align}
For this Ansatz with $m\neq0$, this degeneracy condition takes the explicit form
\begin{align}
	\label{eq:DegeneracyCondition}
	C(\theta)Z(\theta)+C(\theta)Y'(\theta)-C'(\theta)Y(\theta)=0.
\end{align}
Schematically, this condition may be satisfied in one of three ways:
\begin{itemize}
\item
$C(\theta)=0$ -- this choice entirely kills off the electric field and corresponds to the Magnetic Type solutions we will derive in section \ref{sec:TypeM}.
\item
$Y(\theta)=Z(\theta)=0$ -- this choice entirely kills off the magnetic field and corresponds to the Electric Type solutions we will derive in section \ref{sec:TypeE}.
\item
$C(\theta)\neq0$ and $Y(\theta)\neq0$ with
\begin{align}
	Z(\theta)=Y(\theta)\frac{C(\theta)}{C'(\theta)}-Y'(\theta),
\end{align}
which leads to a solution with both electric and magnetic fields turned on, corresponding to the Electromagnetic Type solutions we will derive in section \ref{sec:TypeEM}.
\end{itemize}
Intuitively, the degeneracy condition $F\wedge F=0$ states that the electric and magnetic fields are orthogonal, $\vec{E}\cdot\vec{B}=0$. As such, this precursor to the force-free equations may be satisfied by either setting $\vec{E}=0$, or $\vec{B}=0$, or $\vec{E}\perp\vec{B}$, thereby selecting one of the three cases listed above.

\section{Derivation of Magnetic Type solutions}
\label{sec:TypeM}

NHEK has an $\SL(2,\R)\times\U(1)$ isometry group. It is natural to classify solutions to the force-free equations by their transformation properties under these symmetries.

\subsection{$\SL(2,\R)\times\U(1)$-invariant solution}

The simplest possibility consists of a solution that is invariant under the action of $\SL(2,\R)\times\U(1)$. That is, suppose that in addition to the force-free equations \eqref{eq:FFE1}--\eqref{eq:FFE3}, the 2-form $F$ also satisfies
\begin{align}
	\L_{H_0}F=\L_{H_\pm}F=\L_{W_0}F=0.
\end{align}
There is only one such solution that is nonsingular and has $\J_\theta=0$. We defer its examination for the moment (as it will form the starting point of our later derivation of the Electric Type solutions in section \ref{sec:TypeE}) and consider instead the gauge potential\footnote{This gauge potential satisfies $\L_{H_0}F=\L_{H_+}F=\L_{W_0}F=0$ but is not invariant under $H_-$ since $\L_{H_-}F\neq0$.}
\begin{align}
	A_0^M(S_0)=S_0\ed\phi,
\end{align}
which is fixed up to an arbitrary function $S_0(\theta)$. As we will prove in section \ref{subsec:ProofTypeM}, this Ansatz solves the force-free equations \eqref{eq:FFE1}--\eqref{eq:FFE3} provided that $S_0(\theta)$ satisfies the following second-order ODE:
\begin{align}
	S_0''-\frac{\Lambda'}{\Lambda}\pa{\frac{1+\Lambda^2}{1-\Lambda^2}}S_0'=0.
\end{align}
This equation has two solutions, one of which is badly singular in the range $\theta\in[0,\pi]$. The other solution is a constant and leads to a trivial 2-form $F=0$. We may express this gauge potential in terms of the isometries as
\begin{align}
	A_0^M(S_0)=\frac{S_0}{2J\Gamma}\br{\frac{\Phi}{\sqrt{2}}H_+-\pa{1-\frac{1}{\Lambda^2}}W_0},
\end{align}
where we introduced a scalar function
\begin{align}
	\Phi(r)=\frac{1}{r}.
\end{align}
Note that $\Phi^h(r)$ is $\U(1)$-invariant and forms a scalar highest-weight representation of $\SL(2,\R)$ with highest weight $h$:
\begin{align}
	\L_{H_+}\Phi^h&=0,\\
	\L_{H_0}\Phi^h&=\Phi^h,\\
	\L_{W_0}\Phi^h&=0.
\end{align}

\subsection{$\SL(2,\R)$ highest-weight $\U(1)$-invariant solution}

We now relax the requirement of $\SL(2,\R)$-invariance. Instead, it is natural to search for solutions that form highest-weight representations of $\SL(2,\R)$ with some highest weight $h$. That is, assume that in addition to the force-free equations \eqref{eq:FFE1}--\eqref{eq:FFE3}, the 2-form $F$ also satisfies the conditions
\begin{align}
	\label{eq:HWI1}
	\L_{H_+}F&=0,\\
	\label{eq:HWI2}
	\L_{H_0}F&=hF,\\
	\label{eq:HWI3}
	\L_{W_0}F&=0,
\end{align}
for some $h\in\R$. Consider as an Ansatz the gauge potential
\begin{align}
	A_h^M(S_h)=XA_0^M(S_h),
\end{align}
whose dependence on the coordinates has been split between two arbitrary functions $X(t,r,\phi)$ and $S_h(\theta)$. Imposing the 3 conditions \eqref{eq:HWI1}--\eqref{eq:HWI3} entirely determines the scalar prefactor $X$ to be
\begin{align}
	X(t,r,\phi)=\Phi^h(r).
\end{align}
Our Ansatz for the gauge potential then becomes
\begin{align}
	A_h^M(S_h)&=\Phi^hS_h\ed\phi=\frac{S_h}{r^h}\ed\phi\\
	&=\frac{\Phi^hS_h}{2J\Gamma}\br{\frac{\Phi}{\sqrt{2}}H_+-\pa{1-\frac{1}{\Lambda^2}}W_0}.
\end{align}
As we will prove in section \ref{subsec:ProofTypeM}, this Ansatz solves the force-free equations \eqref{eq:FFE1}--\eqref{eq:FFE3} provided that $S_h(\theta)$ satisfies the following second-order ODE:
\begin{align}
	S_h''-\frac{\Lambda'}{\Lambda}\pa{\frac{1+\Lambda^2}{1-\Lambda^2}}S_h'+h(h-1)S_h=0.
\end{align}
This equation is manifestly invariant under $h\to1-h$, so $S_h(\theta)=S_{1-h}(\theta)$. For the special values $h=0$ and $h=1$, this equation for $S_h(\theta)$ admits a constant solution and the gauge field becomes $\theta$-independent. Observe also when $h=0$, it reduces to the $\SL(2,\R)\times\U(1)$-invariant case considered in the previous section. The solution $S_h(\theta)$ is given in appendix \ref{appendix:ThetaAnalysis}.

\subsection{$\SL(2,\R)$ highest-weight $\U(1)$-eigenstate solution}

We next relax the requirement of $\U(1)$-invariance by allowing the gauge field to carry a $\U(1)$ charge. Suppose that in addition to the force-free equations \eqref{eq:FFE1}--\eqref{eq:FFE3}, the 2-form $F$ also satisfies
\begin{align}
	\L_{H_+}F&=0,\\
	\L_{H_0}F&=hF,\\
	\L_{W_0}F&=imF,
\end{align}
where $h,m\in\R$. As before, we could use as an Ansatz the $\SL(2,\R)\times\U(1)$-invariant gauge potential with a yet-to-be-determined scalar prefactor $\tilde{X}(t,r,\phi)$:
\begin{align}
	A_{h,m}^M(S_{h,m})=\tilde{X}A_0^M(S_{h,m}).
\end{align}
However, it is not possible to satisfy all our conditions with a gauge field of this form, and an additional term dependent on $\Theta=2J\Gamma\ed\theta$ is needed. Instead, we consider the Ansatz
\begin{align}
	A_{h,m}^M(S_{h,m})=\tilde{X}\br{A_0^M(S_{h,m})+\tilde{S}_{h,m}\Theta},
\end{align}
whose dependence on the coordinates has been split between an arbitrary function $\tilde{X}(t,r,\phi)$ and two arbitrary functions $S_{h,m}(\theta)$ and $\tilde{S}_{h,m}(\theta)$.  Imposing the 3 conditions \eqref{eq:HW1}--\eqref{eq:HW3} entirely determines the scalar prefactor $\tilde{X}$ to be\begin{align}
	\tilde{X}(t,r,\phi)&=\Phi^h(r)\Psi^m(\phi),
\end{align}
where we introduced a new scalar function
\begin{align}
	\Psi(\phi)=e^{i\phi}.
\end{align}
Note that $\Psi^m(\phi)$ is $\SL(2,\R)$-invariant and a $\U(1)$-eigenstate:
\begin{align}
	\L_{H_+}\Psi^m&=0,\\
	\L_{H_0}\Psi^m&=0,\\
	\L_{W_0}\Psi^m&=im\Psi.
\end{align}
As we will prove in section \ref{subsec:ProofTypeM}, this Ansatz solves the force-free equations \eqref{eq:FFE1}--\eqref{eq:FFE3} provided that we define
\begin{align}
	\tilde{S}_{h,m}=-im\chi_{h,m}\pa{1-\frac{1}{\Lambda^2}}\frac{S_{h,m}'}{S_{h,m}},
\end{align}
where $\chi_{h,m}(\theta)$ is manifestly invariant under $h\to1-h$ and $m\to-m$,
\begin{align}
	\label{eq:Chi}
	\chi_{h,m}=\br{h(h-1)+m^2\pa{1-\frac{1}{\Lambda^2}}}^{-1},
\end{align}
and that $S_{h,m}(\theta)$ satisfies the following second-order ODE:
\begin{align}
	\label{eq:ThetaTypeM}
	S_{h,m}''+\frac{\Lambda'}{\Lambda}\br{1-\frac{2h(h-1)\chi_{h,m}}{1-\Lambda^2}}S_{h,m}'
	+\frac{S_{h,m}}{\chi_{h,m}}=0.
\end{align}
This equation has the same symmetries as $\chi_{h,m}(\theta)$ -- in particular $S_{h,m}(\theta)=S_{1-h,m}(\theta)$. The solution $S_{h,m}(\theta)$ is given in appendix \ref{appendix:ThetaAnalysis}. The gauge potential is thus
\begin{align}
	A_{h,m}^M(S_{h,m})
	&=\Phi^h\Psi^m\br{S_{h,m}\ed\phi-im\chi_{h,m}\pa{1-\frac{1}{\Lambda^2}}S_{h,m}'\ed\theta}\\
	&=\frac{\Phi^h\Psi^mS_{h,m}}{2J\Gamma}\br{\frac{\Phi}{\sqrt{2}}H_+
		-\pa{1-\frac{1}{\Lambda^2}}\pa{W_0+im\chi_{h,m}\frac{S_{h,m}'}{S_{h,m}}\Theta}}.
\end{align}
When $m=0$, $A_{h,m}^M(S_{h,m})$ reduces to the $\U(1)$-invariant solution $A_h^M(S_h)$ discussed previously.

\subsection{Verification of Magnetic Type solutions}
\label{subsec:ProofTypeM}

We now wish to prove that the Magnetic Type solutions examined above do indeed solve the force-free equations \eqref{eq:FFE1}--\eqref{eq:FFE3}. It suffices to check that $A_{h,m}^M(S_{h,m})$ is a solution. The corresponding 2-form gauge connection $F_{h,m}^M=\ed A_{h,m}^M$ is
\begin{align}
	F_{h,m}^M=-h\Phi^h\Psi^m\cu{\Phi S_{h,m}\ed r\wedge\ed\phi-\chi_{h,m}S_{h,m}'
		\br{im\Phi\pa{1-\frac{1}{\Lambda^2}}\ed r\wedge\ed\theta+(h-1)\ed\theta\wedge\ed\phi}}.
\end{align}
One may then check that
\begin{align}
	\label{eq:TypeMstarF}
	\star F_{h,m}^M=-\frac{h}{\sqrt{2}}\frac{\Phi^{h+1}\Psi^m}{(2J\Gamma)^2\Lambda}\cu{
		S_{h,m}H_+\wedge\Theta+\chi_{h,m}S_{h,m}'\br{
		im\pa{1-\frac{1}{\Lambda^2}}H_+\wedge W_0-(h-1)H_+\wedge H_0}}.
\end{align}
The associated current $\J_{h,m}^M=\ed^\dagger F_{h,m}^M$ is proportional to $H_+$:
\begin{align}
	\label{eq:TypeMCurrent}
	\J_{h,m}^M=\frac{h}{\sqrt{2}}\frac{\Phi^{h+1}\Psi^m}{(2J\Gamma)^2}\br{
		S_{h,m}+\frac{2(h-1)\chi_{h,m}}{1-\Lambda^2}\frac{\Lambda'}{\Lambda}S_{h,m}'}H_+.
\end{align}
This current is only null when $\Lambda(\theta)=1$, corresponding to the boundary of the ergosphere \eqref{eq:Ergosphere}:
\begin{align}
	\pa{\J_{h,m}^M}^2=h^2\frac{\pa{\Phi^h\Psi^m\Lambda}^2}{(2J\Gamma)^3}
	\pa{1-\frac{1}{\Lambda^2}}
	\br{S_{h,m}+\frac{2(h-1)\chi_{h,m}}{1-\Lambda^2}\frac{\Lambda'}{\Lambda}S_{h,m}'}^2.
\end{align}
Since $H_+\wedge H_+=0$, it is evident that
\begin{align}
	\star F_{h,m}^M\wedge\J_{h,m}^M=0,
\end{align}
which proves that this is indeed a solution to  \eqref{eq:FFE1}--\eqref{eq:FFE3}. Note that for this (complex) solution, $F^2>0$ outside the ergosphere, where $\Lambda^2<1$:
\begin{align}
	F^2=-2h^2\pa{\frac{\Phi^h\Psi^m}{2J\Gamma}}^2\pa{1-\frac{1}{\Lambda^2}}\cu{
		S_{h,m}^2+\br{(h-1)^2+m^2\pa{1-\frac{1}{\Lambda^2}}}\chi_{h,m}^2S_{h,m}'^2}.
\end{align}

\section{Derivation of Electric Type solutions}
\label{sec:TypeE}

We now turn to the derivation of the Electric Type solutions. Following the same approach as in the previous section, we attempt to classify solutions to the force-free equations by their transformation properties under the $\SL(2,\R)\times\U(1)$ isometry group of NHEK.

\subsection{$\SL(2,\R)\times\U(1)$-invariant solution}

As before, we start with the simplest possibility, which consists of a solution that is invariant under the action of $\SL(2,\R)\times\U(1)$. That is, suppose that in addition to the force-free equations \eqref{eq:FFE1}--\eqref{eq:FFE3}, the 2-form $F$ also satisfies
\begin{align}
	\L_{H_0}F=\L_{H_\pm}F=\L_{W_0}F=0.
\end{align}
The only such solution that is nonsingular and has $\J_\theta=0$ is obtained from the Ansatz
\begin{align}
	A_0^E(S_0)=-rP_0\ed t,
\end{align}
which is fixed up to an arbitrary function $P_0(\theta)$. As we will prove in section \ref{subsec:ProofTypeE}, this Ansatz solves the force-free equations \eqref{eq:FFE1}--\eqref{eq:FFE3} provided that $P_0(\theta)$ satisfies the following second-order ODE:
\begin{align}
	P_0''+\frac{\Lambda'}{\Lambda}P_0'=0.
\end{align}
This equation has two solutions, one of which diverges at the poles $\theta\in\cu{0,\pi}$. The other solution is a constant, and it leads to an $\SL(2,\R)\times\U(1)$-invariant solution to the force-free equations with $\J\neq0$. We may express this gauge potential in terms of the isometries and the scalar $\Phi$ as
\begin{align}
	A_0^E(P_0)=\frac{P_0}{2J\Gamma}\pa{\frac{\Phi}{\sqrt{2}}H_+-W_0}.
\end{align}

\subsection{$\SL(2,\R)$ highest-weight $\U(1)$-invariant solution}

We now relax the requirement of $\SL(2,\R)$-invariance and search for solutions that form highest-weight representations of $\SL(2,\R)$ with some highest weight $h$. That is, assume that in addition to the force-free equations \eqref{eq:FFE1}--\eqref{eq:FFE3}, the 2-form $F$ also satisfies the conditions \eqref{eq:HW1}--\eqref{eq:HW3} for some $h\in\R$. Consider as an Ansatz the gauge potential
\begin{align}
	A_h^E(P_h)=XA_0^E(P_h),
\end{align}
whose dependence on the coordinates has again been split between two arbitrary function $X(t,r,\phi)$ and $P_h(\theta)$. As before, imposing the 3 conditions \eqref{eq:HW1}--\eqref{eq:HW3} entirely determines the scalar prefactor $X$ to be
\begin{align}
	X(t,r,\phi)=\Phi^h(r).
\end{align}
Our Ansatz for the gauge potential then becomes
\begin{align}
	A_h^E(P_h)&=-\Phi^{h-1}P_h\ed t=-\frac{P_h}{r^{h-1}}\ed t\\
	&=\frac{\Phi^hP_h}{2J\Gamma}\pa{\frac{\Phi}{\sqrt{2}}H_+-W_0}.
\end{align}
As we will prove in section \ref{subsec:ProofTypeE}, this Ansatz solves the force-free equations \eqref{eq:FFE1}--\eqref{eq:FFE3} provided that $P_h(\theta)$ satisfies the following second-order ODE:
\begin{align}
	\label{eq:ThetaTypeE}
	P_h''+\frac{\Lambda'}{\Lambda}P_h'+h(h-1)P_h=0.
\end{align}
This equation is manifestly invariant under $h\to1-h$, so $P_h(\theta)=P_{1-h}(\theta)$. For the special values $h=0$ and $h=1$, this equation for $P_h(\theta)$ admits a constant solution and the gauge field becomes $\theta$-independent. Observe also when $h=0$, it reduces to the $\SL(2,\R)\times\U(1)$-invariant case considered in the previous section. The solution $P_h(\theta)$ is given in appendix \ref{appendix:ThetaAnalysis}.

\subsection{Verification of Electric Type $\U(1)$-invariant solutions}
\label{subsec:ProofTypeE}

We now wish to prove that the Electric Type solutions examined above do indeed solve the force-free equations \eqref{eq:FFE1}--\eqref{eq:FFE3}. It suffices to check that $A_h^E(P_h)$ is a solution. The corresponding 2-form gauge connection $F_h^E=\ed A_h^E$ is
\begin{align}
	F_h^E=-\Phi^{h-1}\br{(h-1)\Phi P_h\ed t\wedge\ed r-P_h'\ed t\wedge\ed\theta}.
\end{align}
One may then check that
\begin{align}
	\star F_h^E=-\frac{\Phi^h}{(2J\Gamma)^2\Lambda}\br{
		(h-1)P_hW_0\wedge\Theta-P_h'\pa{\frac{t}{\sqrt{2}}H_+-H_0}\wedge W_0}.
\end{align}
The associated current $\J_h^E=\ed^\dagger F_h^E$ is
\begin{align}
	\J_h^E=(h-1)\frac{\Phi^hP_h}{(2J\Gamma)^2}W_0.
\end{align}
This current is always spacelike:
\begin{align}
	\pa{\J_h^E}^2=(h-1)^2\frac{\pa{\Phi^h\Lambda P_h}^2}{(2J\Gamma)^3}>0.
\end{align}
Since $W_0\wedge W_0=0$, it is evident that
\begin{align}
	\star F_h^E\wedge\J_h^E=0,
\end{align}
which proves that this is indeed a solution to  \eqref{eq:FFE1}--\eqref{eq:FFE3}. Note that for this (complex) solution, $F^2<0$ everywhere, as expected of a purely electric solution:
\begin{align}
	F^2=-2\pa{\frac{\Phi^h}{2J\Gamma}}^2\br{(h-1)^2P_h^2+P_h'^2}<0.
\end{align}

\subsection{$\SL(2,\R)$ highest-weight $\U(1)$-eigenstate solution}

We next relax the requirement of $\U(1)$-invariance by allowing the gauge field to carry a $\U(1)$ charge. That is, suppose that in addition to the force-free equations \eqref{eq:FFE1}--\eqref{eq:FFE3}, the 2-form $F$ also satisfies \eqref{eq:HW1}--\eqref{eq:HW3} for some $h,m\in\R$. Consider as an Ansatz the gauge potential
\begin{align}
	A_{h,m}^E(P_{h,m})=\tilde{X}A_0^E(P_{h,m}),
\end{align}
whose dependence on the coordinates has again been split between two arbitrary functions $\tilde{X}(t,r,\phi)$ and $P_{h,m}(\theta)$. As before, imposing the 3 conditions \eqref{eq:HW1}--\eqref{eq:HW3} entirely determines the scalar prefactor $\tilde{X}$ to be
\begin{align}
	\tilde{X}(t,r,\phi)&=\Phi^h(r)\Psi^m(\phi).
\end{align}
Our Ansatz for the gauge potential then becomes
\begin{align}
	A_{h,m}^E(P_{h,m})&=-\Phi^{h-1}\Psi^mP_{h,m}\ed t=-\frac{e^{im\phi}P_{h,m}}{r^{h-1}}\ed t\\
	&=\frac{\Phi^h\Psi^mP_{h,m}}{2J\Gamma}\pa{\frac{\Phi}{\sqrt{2}}H_+-W_0}.
\end{align}
It turns out that this Ansatz leads to a solution only when $h\in\cu{0,1}$, as one may explicitly check by first solving for the $t$ component of the force-free condition \eqref{eq:FFE3}. In either case, we may only obtain a solution to the force-free equations \eqref{eq:FFE1}--\eqref{eq:FFE3} by setting
\begin{align}
	P_{0,m}(\theta)=P_{1,m}(\theta)=P_m(\theta),
\end{align}
where
\begin{align}
	\label{eq:ThetaTypeEcharged}
	P_m(\theta)=\int\!\ed\theta\,e^{m/\Lambda(\theta)}
	=C\cosh\pa{\frac{m}{2}\cos\theta+\log\tan^m\frac{\theta}{2}}
	+D\sinh\pa{\frac{m}{2}\cos\theta+\log\tan^m\frac{\theta}{2}}.
\end{align}

\subsubsection{$h=1$ case}

In this particular case, the gauge potential reduces to
\begin{align}
	A_{1,m}^E(P_m)=\frac{\Phi\Psi^mP_m}{2J\Gamma}\pa{\frac{\Phi}{\sqrt{2}}H_+-W_0}
	=-\Psi^mP_m\ed t=-e^{im\phi}P_m\ed t.
\end{align}
The corresponding 2-form gauge connection $F_{1,m}^E=\ed A_{1,m}^E$ is
\begin{align}
	F_{1,m}^E=\Psi^m\br{imP_m\ed t\wedge\ed\phi+P_m'\ed t\wedge\ed\theta}.
\end{align}
One may then check that
\begin{align}
	\star F_{1,m}^E=\frac{\Phi\Psi^m}{(2J\Gamma)^2\Lambda}\br{
		\pa{H_0-\frac{t}{\sqrt{2}}H_+}\wedge\pa{imP_m\Theta-P_m'W_0}}.
\end{align}
Using the explicit formula \eqref{eq:ThetaTypeEcharged}, this further simplifies to
\begin{align}
	\star F_{1,m}^E=\frac{\Phi\Psi^mP_m}{(2J\Gamma)^2\Lambda}\br{
		\frac{m}{\Lambda}\pa{H_0-\frac{t}{\sqrt{2}}H_+}\wedge\pa{i\Lambda\Theta-W_0}}.
\end{align}
Likewise, using \eqref{eq:ThetaTypeEcharged}, the associated current $\J_{1,m}^E=\ed^\dagger F_{1,m}^E$ is
\begin{align}
	\J_{1,m}^E=m^2\frac{\Phi\Psi^mP_m}{(2J\Gamma\Lambda)^2}\pa{i\Lambda\Theta-W_0},
\end{align}
which is null: $\pa{\J_{1,m}^E}^2=0$. Since $\pa{i\Lambda\Theta-W_0}\wedge\pa{i\Lambda\Theta-W_0}=0$, it is evident that
\begin{align}
	\star F_{1,m}^E\wedge\J_{1,m}^E=0,
\end{align}
which proves that this is indeed a solution to \eqref{eq:FFE1}--\eqref{eq:FFE3}. Note that for this (complex) solution, $F^2=0$ everywhere -- this is related to the condition that the current is null.

\subsubsection{$h=0$ case}

In this particular case, the gauge potential reduces to
\begin{align}
	A_{0,m}^E(P_m)=\frac{\Psi^mP_m}{2J\Gamma}\pa{\frac{\Phi}{\sqrt{2}}H_+-W_0}
	=-\Phi^{h-1}\Psi^mP_{h,m}\ed t=-\frac{e^{im\phi}P_{h,m}}{r^{h-1}}\ed t.
\end{align}
The corresponding 2-form gauge connection $F_{0,m}^E=\ed A_{0,m}^E$ is
\begin{align}
	F_{0,m}^E=\frac{\Psi^m}{\Phi}\br{
		\Phi P_m\ed t\wedge\ed r+imP_m\ed t\wedge\ed\phi+P_m'\ed t\wedge\ed\theta}.
\end{align}
One may then check that
\begin{align}
	\star F_{0,m}^E=\frac{\Psi^m}{(2J\Gamma)^2\Lambda}\br{
		\pa{H_0-\frac{t}{\sqrt{2}}H_+}\wedge\pa{imP_m\Theta-P_m'W_0}+P_mW_0\wedge\Theta}.
\end{align}
Using the explicit formula \eqref{eq:ThetaTypeEcharged}, this further simplifies to
\begin{align}
	\star F_{0,m}^E=\frac{\Psi^mP_m}{(2J\Gamma)^2\Lambda}\br{
		\frac{m}{\Lambda}\pa{H_0-\frac{t}{\sqrt{2}}H_+}\wedge\pa{i\Lambda\Theta-W_0}+W_0\wedge\Theta}.
\end{align}
Likewise, using \eqref{eq:ThetaTypeEcharged}, the associated current $\J_{0,m}^E=\ed^\dagger F_{0,m}^E$ is
\begin{align}
	\J_{0,m}^E=\frac{\Psi^mP_m}{(2J\Gamma\Lambda)^2}\br{
		m^2\pa{i\Lambda\Theta-W_0}-im\Lambda^2\pa{H_0-\frac{t}{\sqrt{2}}H_+}-\Lambda^2W_0},
\end{align}
which is always spacelike:
\begin{align}
	\pa{\J_{0,m}^E}^2=\frac{\pa{\Psi^mP_m}^2}{(2J\Gamma)^3}\pa{m^2+\Lambda^2}>0.
\end{align}
Since $\omega\wedge\omega=0$ for any form $\omega$, it is evident that
\begin{align}
	\star F_{0,m}^E\wedge\J_{0,m}^E
	&=\frac{\Psi^mP_m}{(2J\Gamma)^2\Lambda}
		\br{\frac{m}{\Lambda}\pa{H_0-\frac{t}{\sqrt{2}}H_+}\wedge\pa{i\Lambda\Theta}}
		\wedge\br{-\frac{\Psi^m}{(2J\Gamma)^2}P_mW_0}\nonumber\\
	&\quad+\br{\frac{\Psi^mP_m}{(2J\Gamma)^2\Lambda}W_0\wedge\Theta}
		\wedge\frac{\Psi^mP_m}{(2J\Gamma\Lambda)^2}\br{-im\Lambda^2\pa{H_0-\frac{t}{\sqrt{2}}H_+}}\\
	&=0,\nonumber
\end{align}
which proves that this is indeed a solution to \eqref{eq:FFE1}--\eqref{eq:FFE3}. Note that for this (complex) solution, $F^2<0$ everywhere, as expected of a purely electric solution:
\begin{align}
	F^2=-2\pa{\frac{\Psi^mP_m}{2J\Gamma}}^2<0.
\end{align}

\section{Derivation of Electromagnetic Type solutions}
\label{sec:TypeEM}

We now turn to the derivation of the Electromagnetic Type solutions, which also share the same transformation properties under the $\SL(2,\R)\times\U(1)$ isometry group of NHEK.

It is quite difficult to obtain these solutions directly. However, one may construct them as nonlinear superpositions of the purely electric and purely magnetic solutions. This mechanism yields solutions with current flowing in the $\theta$-direction, $\J_\theta\neq0$.

\subsection{$\SL(2,\R)$ highest-weight $\U(1)$-invariant solution}
\label{subsec:AxisymmetricMagnetosphere}

In the foregoing discussion, we derived two $\SL(2,\R)$ highest-weight $\U(1)$-invariant solutions to the force-free equations, one of which was purely magnetic,
\begin{align}
	A_h^M(S_h)=\frac{\Phi^hS_h}{2J\Gamma}\br{\frac{\Phi}{\sqrt{2}}H_+-\pa{1-\frac{1}{\Lambda^2}}W_0},
\end{align}
and another which was purely electric,
\begin{align}
	A_h^E(P_h)=\frac{\Phi^hP_h}{2J\Gamma}\pa{\frac{\Phi}{\sqrt{2}}H_+-W_0}.
\end{align}
In free Maxwell electrodynamics, where the equations are linear, one could then obtain electromagnetic solutions by forming linear combinations of these potentials. However, this cannot work in the context of force-free electrodynamics, where the equations are nonlinear. Nonetheless, we can explore the possibility of nonlinear superpositions. To that end, consider as an Ansatz the potential
\begin{align}
	A_h^{EM}(U_h,V_h)=A_h^M(U_h)+A_h^E(V_h),
\end{align}
where the functions $U_h(\theta),V_h(\theta)$ need not be identical to the functions $P_h(\theta),S_h(\theta)$ introduced in sections \ref{sec:TypeM} and \ref{sec:TypeE}. Since they are (as of now) unconstrained functions of $\theta$, this is not a linear combination, but instead a truly nonlinear superposition (with nonlinearity in $\theta$). More generally, one could consider including an additional factor in the $\theta$-direction; that is, one could Ansatz
\begin{align}
	A_h^{EM}(U_h,V_h,T_h)=A_h^M(U_h)+A_h^E(V_h)+\frac{\Phi^hT_h}{2J\Gamma}\Theta,
\end{align}
where the dependence of the last term on the coordinates $(t,r,\phi)$ is fixed to be $\Phi^h(r)$ in order to ensure that $A_h^{EM}(T_h,U_h,V_h)$ is still an $\SL(2,\R)$ highest-weight $\U(1)$-invariant potential. Explicitly, this corresponds to
\begin{align}
	A_h^{EM}(U_h,V_h,T_h)=-\Phi^{h-1}V_h\ed t+\Phi^hT_h\ed\theta+\Phi^hU_h\ed\phi
	=-\frac{V_h}{r^{h-1}}\ed t+\frac{T_h}{r^h}\ed\theta+\frac{U_h}{r^h}\ed\phi.
\end{align}
The corresponding 2-form gauge connection $F_h^{EM}=\ed A_h^{EM}$ is
\begin{align}
	F_h^{EM}=\Phi^h\br{-(h-1)V_h\ed t\wedge\ed r+\frac{V_h'}{\Phi}\ed t\wedge\ed\theta
		-h\Phi\pa{T_h\ed r\wedge\ed\theta+U_h\ed r\wedge\ed\phi}+U_h'\ed\theta\wedge\ed\phi}.
\end{align}
One may then check that
\begin{align}
	\star F_h^{EM}
	&=\frac{\Phi^h}{(2J\Gamma)^2\Lambda}\br{
		\frac{\Phi}{\sqrt{2}}H_+\wedge\pa{U_h'H_0+hT_hW_0-hU_h\Theta}
		-V_h'	\pa{H_0-\frac{t}{\sqrt{2}}H_+}\wedge W_0-(h-1)V_hW_0\wedge\Theta}.
\end{align}
The associated current $\J_h^{EM}=\ed^\dagger F_h^{EM}$ is
\begin{align}
	\J_h^{EM}=\frac{\Phi^h}{(2J\Gamma)^2\Lambda}\br{
		\frac{\Phi}{\sqrt{2}}\kappa H_+-\lambda W_0
		-h\pd_\theta\pa{\Lambda T_h}\pa{H_0-\frac{t}{\sqrt{2}}H_+}
		+h(h-1)\Lambda T_h\Theta},
\end{align}
where $\kappa(\theta)$ and $\lambda(\theta)$ are given by
\begin{align}
	\kappa&=\pd_\theta\br{\Lambda\pa{U_h'+V_h'}}+h\Lambda\br{hU_h+(h-1)V_h},\\
	\lambda&=\pd_\theta\cu{\Lambda(\theta)\br{U_h'\pa{1-\frac{1}{\Lambda^2}}+V_h'}}
	+h(h-1)\Lambda U_h\pa{1-\frac{1}{\Lambda^2}}+(h-1)^2\Lambda V_h.
\end{align}
This current is not null:
\begin{align}
	\pa{\J_h^{EM}}^2=\frac{\Phi^{2h}}{(2J\Gamma)^3}\br{
		h^2(h-1)^2T_h^2+h^2T_h'^2+h^2\frac{\Lambda'}{\Lambda^2}\pd_\theta\pa{\Lambda T_h^2}
		+\pa{\kappa-\lambda}^2-\pa{\frac{\kappa}{\Lambda}}^2}.
\end{align}
Note that for this (real) solution, $F^2\neq0$:
\begin{align}
	F^2=2\pa{\frac{\Phi^h}{2J\Gamma}}^2\cu{\frac{U_h'^2+h^2U_h^2}{\Lambda^2}
		-\pa{U_h'+V_h'}^2-\br{hU_h+(h-1)V_h}^2+h^2T_h^2}.
\end{align}
For this to be a solution to the force-free equations, we must now impose the force-free condition
\begin{align}
	\star F_h^{EM}\wedge\J_h^{EM}=0.
\end{align}
This relation implies the weaker degeneracy condition $F\wedge F=0$, from which we immediately see that we must require
\begin{align}
	\label{eq:V}
	V_h(\theta)=\frac{Dh}{h-1}U_h(\theta)^{1-1/h},
\end{align}
for some constant $D$. (It is especially nice that we can eliminate $V_h(\theta)$ from a purely algebraic condition.) For the $t$ component of the force-free equations to be satisfied (in order to ensure $F_{\mu t}\J^\mu=0$), we must then demand that
\begin{align}
	\label{eq:T}
	T_h(\theta)=\frac{C}{\Lambda(\theta)}U_h(\theta)^{1-1/h},
\end{align}
for some constant $C$. Using \eqref{eq:V} and \eqref{eq:T}, the force-free equations \eqref{eq:FFE1}--\eqref{eq:FFE3} may be satisfied provided that $U_h(\theta)$ satisfies the following second-order nonlinear ODE:
\begin{align}
	\label{eq:ThetaTypeEM}
	&\frac{1}{\Lambda U_h}\br{\pa{D+U_h^{1/h}}^2\pd_\theta\pa{\Lambda U_h'}
		-U_h^{2/h}\pd_\theta\pa{\frac{U_h'}{\Lambda}}}\\
	&+\pa{D+U_h^{1/h}}\br{Dh^2+h(h-1)U_h^{1/h}-\frac{D}{h}\pa{\frac{U_h'}{U_h}}^2}
	-\frac{h(h-1)}{\Lambda^2}\pa{C^2+U_h^{2/h}}=0.\nonumber
\end{align}
The solution can be succinctly written as
\begin{align}
	\label{eq:TypeEM}
	A_h^{EM}(U_h)	&=\frac{\Phi^hU_h}{2J\Gamma}\br{\frac{\Phi}{\sqrt{2}}\pa{1+\frac{DhU_h^{-1/h}}{h-1}}H_+
		-\pa{1-\frac{1}{\Lambda^2}+\frac{DhU_h^{-1/h}}{h-1}}W_0+\frac{CU_h^{-1/h}}{\Lambda}\Theta},
\end{align}
with $U_h$ defined by \eqref{eq:ThetaTypeEM}. This is likely related to the solution recently published in \cite{Zhang:2014pla}.

\subsection{$\SL(2,\R)$ highest-weight $\U(1)$-eigenstate solution}

In the last section, we nonlinearly combined the $\U(1)$-invariant solutions of Electric Type and Magnetic Type to generate a $\U(1)$-invariant solution of Electromagnetic Type. It is natural to ask whether this approach still works in the presence of $\U(1)$-charge or, equivalently, when $m\neq0$.

Recall that we derived two $\SL(2,\R)$ highest-weight $\U(1)$-eigenstate solutions to the force-free equations, one of which was purely magnetic,
\begin{align}
	A_{h,m}^M(S_{h,m})=\frac{\Phi^h\Psi^mS_{h,m}}{2J\Gamma}\br{\frac{\Phi}{\sqrt{2}}H_+
		-\pa{1-\frac{1}{\Lambda^2}}\pa{W_0+im\chi_{h,m}\frac{S_{h,m}'}{S_{h,m}}\Theta}},
\end{align}
and another which was purely electric,
\begin{align}
	A_{h,m}^E(P_{h,m})=\frac{\Phi^h\Psi^mP_{h,m}}{2J\Gamma}\pa{\frac{\Phi}{\sqrt{2}}H_+-W_0},
\end{align}
but only valid for $h\in\cu{0,1}$. Proceeding by analogy with the $m=0$ case, we now Ansatz
\begin{align}
	A_{h,m}^{EM}(U_{h,m},V_{h,m},T_{h,m})
	=A_{h,m}^M(U_{h,m})+A_{h,m}^E(V_{h,m})+\frac{\Phi^h\Psi^mT_{h,m}}{2J\Gamma}\Theta,
\end{align}
where the dependence of the additional $\Theta$ factor on the coordinates $(t,r,\phi)$ is fixed to be $\Phi^h(r)\Psi^m(\phi)$ in order to ensure that the potential $A_{h,m}^{EM}(U_{h,m},V_{h,m},T_{h,m})$ is still an $\SL(2,\R)$ highest-weight $\U(1)$-eigenstate. Again, note that this is a nonlinear superposition in $\theta$ because the functions $U_{h,m}(\theta),V_{h,m}(\theta)$ need not be identical to the functions $P_{h,m}(\theta),S_{h,m}(\theta)$ introduced in sections \ref{sec:TypeM} and \ref{sec:TypeE}. Following the same procedure as before, we obtain the Ansatz
\begin{align}
	A_{h,m}^{EM}(U_{h,m})&=\Phi^{h-1}\Psi^m\br{-h(h-1)U_{h,m}\ed t
		+m^2\Phi U_{h,m}\ed\phi-im\Phi U_{h,m}'\ed\theta}\\
	&=\frac{\Phi^h\Psi^mU_{h,m}}{2J\Gamma}\cu{
		\frac{\Phi}{\sqrt{2}}\br{h(h-1)+m^2}H_+
		-\frac{W_0}{\chi_{h,m}}-im\frac{U_{h,m}'}{U_{h,m}}\Theta},
\end{align}
with $\chi_{h,m}(\theta)$ as in \eqref{eq:Chi}. As we will prove momentarily, this Ansatz solves the force-free equations \eqref{eq:FFE1}--\eqref{eq:FFE3} provided that $U_{h,m}(\theta)$ satisfies the following second-order ODE:
\begin{align}
	U_{h,m}''+\frac{\Lambda'}{\Lambda}U_{h,m}'+\frac{U_{h,m}}{\chi_{h,m}}=0.
\end{align}
This equation has the same symmetries as $\chi_{h,m}(\theta)$: $U_{h,m}(\theta)=U_{1-h,m}(\theta)$ and $U_{h,m}(\theta)=U_{h,-m}(\theta)$.

When $m=0$, $A_{h,m}^{EM}(U_{h,m})$ reduces to the purely electric $\U(1)$-invariant solution $A_h^E(S_h)$ discussed previously, but when the $\U(1)$ charge is non-vanishing ($m\neq0$), the solution is no longer strictly electric. The corresponding 2-form gauge connection $F_{h,m}^{EM}=\ed A_{h,m}^{EM}$ is
\begin{align}
	F_{h,m}^{EM}=-h\Phi^{h-1}\Psi^m\bigg\{&
		(h-1)^2\Phi U_{h,m}\ed t\wedge\ed r-(h-1)U_{h,m}'\ed t\wedge\ed\theta
		-i(h-1)mU_{h,m}\ed t\wedge\ed\phi\nonumber\\
		&+m^2\Phi^2U_{h,m}\ed r\wedge\ed\phi
		-im\Phi^2U_{h,m}'\ed r\wedge\ed\theta
	\bigg\}.
\end{align}
One may then check that
\begin{align}
	\star F_{h,m}^{EM}
	&=h\frac{\Phi^h\Psi^m}{(2J\Gamma)^2\Lambda}\cu{
		\br{(h-1)H_0-\frac{(h-1)t-im\Phi}{\sqrt{2}}H_+}\wedge\pa{imU_{h,m}\Theta-U_{h,m}'W_0}
		-(h-1)^2U_{h,m}W_0\wedge\Theta}.
\end{align}
The associated current $\J_{h,m}^{EM}=\ed^\dagger F_{h,m}^{EM}$ is
\begin{align}
	\J_{h,m}^{EM}=h\frac{\Phi^h\Psi^mU_{h,m}}{(2J\Gamma)^2}\cu{
		\frac{\Phi}{\sqrt{2}}\br{i(h-1)mrt+m^2}H_+-i(h-1)mH_0+(h-1)^2W_0},
\end{align}
which is not null:
\begin{align}
	\pa{\J_{h,m}^{EM}}^2=h^2\frac{\pa{\Phi^h\Psi^m\Lambda U_{h,m}}^2}{(2J\Gamma)^3}
	\br{(h-1)^2+m^2}\br{(h-1)^2+m^2\pa{1-\frac{1}{\Lambda^2}}}.
\end{align}
Though it is not quite obvious, one can explicitly check that
\begin{align}
	\star F_{h,m}^{EM}\wedge\J_{h,m}^{EM}=0,
\end{align}
which proves that this is indeed a solution to \eqref{eq:FFE1}--\eqref{eq:FFE3}. In general, for this (complex) solution $F^2<0$. However, outside the ergoregion and when $m$ is large, it is possible to achieve $F^2>0$:
\begin{align}
	F^2=-2h^2\br{(h-1)^2+m^2}\pa{\frac{\Phi^h\Psi^m}{2J\Gamma}}^2\cu{
		\br{(h-1)^2+m^2\pa{1-\frac{1}{\Lambda^2}}}U_{h,m}^2+U_{h,m}'^2}.
\end{align}

\section{Hidden $\SL(2,\C)$ symmetry}
\label{sec:Automorphisms}

Let us now reexamine one of our solutions, such as the general one of Magnetic Type,
\begin{align}
	A_{h,m}^M(S_{h,m})=\frac{\Phi^h\Psi^mS_{h,m}}{2J\Gamma}\br{\frac{\Phi}{\sqrt{2}}H_+
		-\pa{1-\frac{1}{\Lambda^2}}\pa{W_0+im\chi_{h,m}\frac{S_{h,m}'}{S_{h,m}}\Theta}}.
\end{align}
It is a striking fact that every part of this gauge potential is entirely determined by the Killing vectors $\cu{H_0,H_\pm,W_0}$. In particular, the scalar functions multiplying these vector fields are themselves defined by their transformation properties under the isometries. For instance, $\Phi^h(r)$ is obtained by demanding that it lie in a scalar highest-weight representation of $\SL(2,\R)$ of highest weight $h$.

However, the Killing vectors $\cu{H_0,H_\pm,W_0}$ only form a particular embedding of the $\SL(2,\R)\times\U(1)$ Lie algebra into the NHEK spacetime. This suggests that a different spacetime embedding, which may lead to a different gauge potential, would still produce a force-free solution.

In this section, we will examine all such spacetime embeddings. The ones related to the Poincar\'{e} basis $\cu{H_0,H_\pm,W_0}$ by an $\SL(2,\R)$ transformation will generate physically equivalent solutions obtainable by simple coordinate transformations (finite $\SL(2,\R)$ isometries), but those obtained from the action of $\SL(2,\C)$, the complexification of $\SL(2,\R)$, will produce new solutions. In other words, we will prove the existence of a hidden symmetry of FFE and exploit it to generate new solutions.

\subsection{Automorphism group of $\SL(2,\R)\times\U(1)$}

To explore this idea, we need to find all the possible embeddings of the $\SL(2,\R)\times\U(1)$ Lie algebra into the NHEK spacetime. In other words, we wish to find all the Killing vector fields $\{\tilde{H}_0,\tilde{H}_\pm,\tilde{W}_0\}$ that satisfy the $\SL(2,\R)\times\U(1)$ commutation relations,
\begin{align}
	\br{\tilde{H}_0,\tilde{H}_\pm}=\mp\tilde{H}_\pm,\qquad\br{\tilde{H}_+,\tilde{H}_-}=2\tilde{H}_0,\qquad
	\br{\tilde{W}_0,\tilde{H}_0}=\br{\tilde{W}_0,\tilde{H}_\pm}=0.
\end{align}
We already know one such embedding, $\cu{H_0,H_\pm,W_0}$. All other embeddings are by definition related to it by an automorphism of the $\SL(2,\R)\times\U(1)$ Lie algebra. Since $\SL(2,\R)\times\U(1)$ is its own automorphism group (because it admits no outer automorphisms), these other embeddings can be parameterized by four $\SL(2,\R)\times\U(1)$ parameters $\cu{\alpha,\beta,\gamma,\delta}$ as follows:
\begin{align}
	\tilde{H}_0&=\br{\sqrt{2}\alpha(1+\alpha\beta)+(1+2\alpha\beta)t
		+\frac{\beta}{\sqrt{2}}\pa{t^2+\frac{1}{r^2}}}\pd_t
		-\br{(1+2\alpha\beta)+\sqrt{2}\beta t}r\pd_r-\frac{\sqrt{2}\beta}{r}\pd_\phi,
		\nonumber\\
	\frac{\tilde{H}_+}{\sqrt{2}e^{-\gamma}}&=\br{(1+\alpha\beta)^2+\sqrt{2}\beta(1+\alpha\beta)t
		+\frac{\beta^2}{2}\pa{t^2+\frac{1}{r^2}}}\pd_{t}
		-\beta\br{\sqrt{2}(1+\alpha\beta)+\beta t}r\pd_r-\frac{\beta^2}{r}\pd_\phi,
		\nonumber\\
	\label{eq:GeneralEmbedding}
	\frac{\tilde{H}_-}{\sqrt{2}e^\gamma}&=\br{\alpha^2+\sqrt{2}\alpha t
		+\frac{1}{2}\pa{t^2+\frac{1}{r^2}}}\pd_t-\pa{\sqrt{2}\alpha+t}r\pd_r-\frac{1}{r}\pd_{\phi},\\
	\tilde{W}_0&=\delta\pd_\phi.\nonumber
\end{align}

\subsection{Complex $\SL(2,\C)$ automorphisms as solution-generating mechanism}

In the Poincar\'e basis solutions were obtained by assuming \eqref{eq:HW1}--\eqref{eq:HW3}. One may more generally impose the $(\alpha,\beta)$-dependant conditions
\begin{align}
	\L_{\tilde{H}_+}\tilde{F}&=0,\label{eq:FFEtilde1}\\
	\L_{\tilde{H}_0}\tilde{F}&=h\tilde{F},\label{eq:FFEtilde2}\\
	\L_{\tilde{W}_0}\tilde{F}&=im\tilde{F},\label{eq:FFEtilde3}
\end{align}
The force-free solutions $\tilde{F}=d\tilde{A}$ satisfying these new group-theoretic conditions can be obtained from the old ones by the following $(\alpha,\beta)$-dependent procedure:
\begin{align}
	\label{eq:Procedure}
	A\br{\Phi(r),\Psi(\phi),H_+}&\longrightarrow
	\tilde{A}\br{\tilde{\Phi}(r,t,\alpha,\beta),\tilde{\Psi}(r,t,\phi,\alpha,\beta),e^\gamma\tilde{H}_+},
\end{align}
where the scalar functions $\tilde{\Phi}$ and $\tilde{\Psi}$ are the analogues of $\Phi$ and $\Psi$ with respect to this more general embedding. That is, $\tilde{\Phi}^h\tilde{\Psi}^m$ should be an $\SL(2,\R)$ highest-weight representation with highest weight $h$, and a $\U(1)$-eigenstate with eigenvalue $m$, with respect to the Killing vectors $\cu{\tilde{H}_0,\tilde{H}_\pm,\tilde{W}_0}$:
\begin{align}
	\L_{\tilde{H}_+}\tilde{\Phi}^h\tilde{\Psi}^m&=0,\\
	\L_{\tilde{H}_0}\tilde{\Phi}^h\tilde{\Psi}^m&=h\tilde{\Phi}^h\tilde{\Psi}^m,\\
	\L_{\tilde{W}_0}\tilde{\Phi}^h\tilde{\Psi}^m&=im\tilde{\Phi}^h\tilde{\Psi}^m.
\end{align}
These conditions are satisfied by the $\gamma$-independent functions\footnote{For reasons that will become clear in section \ref{subsec:AutomorphismDetails}, the dependence on $\delta$ eventually drops out. The dependence on $\gamma$ also drops out naturally because $e^\gamma\tilde{H}_+$ is $\gamma$-independent. Thus the $\U(1)$ subgroup of $\SL(2,\R)$ generated by $\gamma$ acts trivially, a point we revisit in greater detail in section \ref{subsec:AutomorphismDetails}.}
\begin{align}
	\label{eq:tildePhi}
	\tilde{\Phi}(t,r,\alpha,\beta)&=\frac{2r}{\br{\sqrt{2}(1+\alpha\beta)+\beta t}^2r^2-\beta^2},\\
	\label{eq:tildePsi}
	\tilde{\Psi}(t,r,\phi,\alpha,\beta,\delta)&=
	\exp\br{C+\frac{i}{\delta}\pa{\phi+2\,\mathrm{arccoth}\frac{\beta}{\br{\sqrt{2}(1+\alpha\beta)+\beta t}r}}},
\end{align}
where setting $C=\pi$ ensures that $\tilde{\Phi}(t,r,0,0)=\Phi(r)$ and $\tilde{\Psi}(t,r,\phi,0,0,1)=\Psi(\phi)$. With this choice of $\tilde{\Phi}$ and $\tilde{\Psi}$, the procedure \eqref{eq:Procedure} generates new solutions to the force-free equations \eqref{eq:FFE1}-\eqref{eq:FFE3}, as described in section \ref{subsec:AutomorphismDetails}.

For real parameters $\alpha,\beta\in\R$ this transformation is equivalent to the coordinate transformation
\begin{align}
	\label{eq:tprime}
	t'&=-\frac{2}{\beta}
	\pa{\frac{\br{\sqrt{2}(1+\alpha\beta)+\beta t}r^2}{\br{\sqrt{2}(1+\alpha\beta)+\beta t}^2r^2-\beta^2}},\\
	\label{eq:rprime}
	r'&=\frac{\br{\sqrt{2}(1+\alpha\beta)+\beta t}^2r^2-\beta^2}{2r},\\
	\phi'&=\phi+2\,\mathrm{arccoth}\frac{\beta}{\br{\sqrt{2}(1+\alpha\beta)+\beta t}r}.
\end{align}
The transformation $\cu{t,r,\theta,\phi}\to\cu{t',r',\theta,\phi'}$ leaves the form of the metric unchanged, and is therefore an isometry\footnote{Observe that \eqref{eq:tildePhi}--\eqref{eq:tildePsi} may also be written as $\tilde{\Phi}(r')=1/r'$ and $\tilde{\Psi}(\phi')=e^{C+i\delta^{-1}\phi'}$.}. To be more exact, it is a ``finite $\SL(2,\R)$ transformation" and as such, it must take solutions of the force-free equations into other (equivalent) solutions.

However, in principle nothing prevents us from allowing instead complex parameters $\alpha,\beta\in\C$. This amounts to considering complex automorphisms of $\SL(2,\R)$ or, more precisely, automorphisms of its complexification $\SL(2,\C)$. The force-free equations \eqref{eq:FFE1}-\eqref{eq:FFE3} will still be satisfied (they are in some sense holomorphic), albeit by complex gauge potentials. These new solutions will be truly different from the original ones, as they will be related by a complex diffeomorphism.

We have thus found a hidden symmetry, stemming from the complex part of $\SL(2,\C)$. To determine what it is, recall that
\begin{align}
	\SL(2,\C)\cong\SL(2,\R)\times\SL(2,\R).
\end{align}
Modding out by the real $\SL(2,\R)$ automorphisms (because they generate physically equivalent solutions) then leaves us with
\begin{align}
	\SL(2,\C)/\SL(2,\R)\cong\SL(2,\R).
\end{align}
However, we should be a bit more careful because, as we will show momentarily, this $\SL(2,\R)$ has a $\U(1)$ subgroup generated by $\gamma$ that induces trivial dilations. When the three parameters $\cu{\alpha,\beta,\gamma}$ of $\SL(2,\R)$ are complexified to six parameters $\cu{\alpha_R,\alpha_I,\beta_R,\beta_I,\gamma_R,\gamma_I}$ of $\SL(2,\C)$, the complex parameter $\gamma_\C=\gamma_R+i\gamma_I$ will still generate a trivial dilation, though this time it will also multiply by a phase in addition to the rescaling factor. As such, we must remove an additional trivial generator, corresponding to the trivial complex $\U(1)$ phase. We are then left with an additional $\SL(2,\R)/\U(1)$ symmetry group.

In conclusion, our fundamental solutions depend on two labels $(h,m)$ that indicate which representation of $\SL(2,\R)\times\U(1)$ they lie in. In addition, they also depend on two parameters $\alpha,\beta\in\C$ that generate an extra $\SL(2,\R)/\U(1)$ symmetry. Therefore, the solutions exhibit an enlarged $\SL(2,\C)$ symmetry group,
\begin{align}
	``\quad\SL(2,\R)\times\U(1)\times\frac{\SL(2,\R)}{\U(1)}\cong\SL(2,\R)\times\SL(2,\R).\quad"
\end{align}

\subsection{Most general highest-weight solutions}
\label{subsec:AutomorphismDetails}

Let us revisit the earlier example of
\begin{align}
	A_{h,m}^M(S_{h,m})=\frac{\Phi^h\Psi^mS_{h,m}}{2J\Gamma}\br{\frac{\Phi}{\sqrt{2}}H_+
		-\pa{1-\frac{1}{\Lambda^2}}\pa{W_0+im\chi_{h,m}\frac{S_{h,m}'}{S_{h,m}}\Theta}}.
\end{align}
We are now ready to perform an $\SL(2,\C)$ automorphism to obtain a new gauge potential
\begin{align}
	\tilde{A}_{h,m}^M(S_{h,m})=\frac{\tilde{\Phi}^h\tilde{\Psi}^mS_{h,m}}{2J\Gamma}
	\br{\frac{\tilde{\Phi}}{\sqrt{2}}\tilde{H}_+
		-\pa{1-\frac{1}{\Lambda^2}}\pa{\tilde{W}_0+im\chi_{h,m}\frac{S_{h,m}'}{S_{h,m}}\tilde{\Theta}}},
\end{align}
which solves the force-free equations \eqref{eq:FFE1}--\eqref{eq:FFE3} provided that $\Theta$ transforms under the $\U(1)$ as
\begin{align}
	\Theta\longrightarrow\tilde{\Theta}=\delta\Theta,
\end{align}
and that we couple the $\SL(2,\R)$ and $\U(1)$ actions by letting
\begin{align}
	\delta=e^{-\gamma}.
\end{align}
But now observe that the transformation generated by $\gamma$ has a trivial action:
\begin{align}
	\tilde{A}_{h,m}^M(S_{h,m})=e^{-\gamma}\br{\tilde{A}_{h,m}^M(S_{h,m})}_{\gamma=0}.
\end{align}
This is just a rescaling of the gauge potential by a constant, and we are free to counter it by multiplying through by $\delta^{-1}=e^\gamma$. This results in
\begin{align}
	\tilde{A}_{h,m}^M(S_{h,m})=\frac{\tilde{\Phi}^h\tilde{\Psi}^mS_{h,m}}{2J\Gamma}
		\br{\frac{\tilde{\Phi}}{\sqrt{2}}\pa{e^\gamma\tilde{H}_+}
		-\pa{1-\frac{1}{\Lambda^2}}\pa{W_0+im\chi_{h,m}\frac{S_{h,m}'}{S_{h,m}}\Theta}}.
\end{align}
We now understand the benefit of working with the $\cu{\alpha,\beta,\gamma}$ parameterization of the $\SL(2,\R)$: it renders manifest the fact that $\SL(2,\R)$ has a $\U(1)$ subgroup (generated by $\gamma$) whose action is a trivial dilation. We are free to ignore it, leaving us with a nontrivial $\SL(2,\R)/\U(1)$ action parameterized by $\alpha$ and $\beta$. It may seem surprising that the $\U(1)$ isometry parameterized by $\delta$ does not produce new solutions and that $\delta$ must instead be related to an $\SL(2,\R)$ parameter. The reason for this is that ``finite $\U(1)$ transformations" are really the angular shifts $\phi\to\phi'=\phi+C$. Observe that $\pd_\phi=\pd_{\phi'}$, so $W_0$ is invariant under this transformation. This symmetry does appear in \eqref{eq:tildePsi}, where it allows us to match $\tilde\Psi$ to $\Psi$ through the normalization factor $C$.

As a final note, we find by the procedure \eqref{eq:Procedure} described in the previous section solutions $\tilde{A}^M$, $\tilde{A}^E$ and $\tilde{A}^{EM}$. These will no longer be purely magnetic, electric or electromagnetic, so instead we shall henceforth refer to them as Type E, Type M and Type EM, respectively. These three families of solutions are labeled by $\SL(2,\C)$ parameters and exhaust the space of highest-weight force-free solutions. They are displayed in Table \ref{table:Classification}.

\begin{table}[t]
\centering
\resizebox{\columnwidth}{!}{
\begin{tabular}{| c | c | c | c |}
\hline
	Vector potential $\Big.\tilde{A}$ & Properties \\
\hline
	& Type M \\
	$\tilde{A}_{0,0}^M(S_0)=\displaystyle
	\frac{S_0}{2J\Gamma}\br{\frac{\tilde{\Phi}}{\sqrt{2}}\tilde{H}_+-\pa{1-\frac{1}{\Lambda^2}}W_0}$ &
	$\SL(2,\R)\times \U(1)$-invariant \\
	& $h=m=0$ \\
\hline
	& Type M \\
	$\tilde{A}_{h,0}^M(S_h)=\displaystyle
	\frac{\tilde{\Phi}^hS_h}{2J\Gamma}\br{\frac{\tilde{\Phi}}{\sqrt{2}}\tilde{H}_+-\pa{1-\frac{1}{\Lambda^2}}W_0}$ &
	$\SL(2,\R)$ highest-weight $\U(1)$-invariant \\
	& $h\in\C$, $m=0$ \\
\hline
	& Type M \\
	$\tilde{A}_{h,m}^M(S_{h,m})=\displaystyle\frac{\tilde{\Phi}^h\tilde{\Psi}^mS_{h,m}}{2J\Gamma}\br{\frac{\tilde{\Phi}}{\sqrt{2}}\tilde{H}_+
		-\pa{1-\frac{1}{\Lambda^2}}\pa{W_0+im\chi_{h,m}\frac{S_{h,m}'}{S_{h,m}}\Theta}}$ &
	$\SL(2,\R)$ highest-weight $\U(1)$-eigenstate \\
	& $h\in\C$, $m\in\Z$ \\
\hline
	& Type E \\
	$\tilde{A}_{0,0}^E(P_0)=\displaystyle
	\frac{P_0}{2J\Gamma}\pa{\frac{\tilde{\Phi}}{\sqrt{2}}\tilde{H}_+-W_0}$ &
	$\SL(2,\R)\times \U(1)$-invariant \\
	& $h=m=0$ \\
\hline
	& Type E \\
	$\tilde{A}_{h,0}^E(P_h)=\displaystyle
	\frac{\tilde{\Phi}^hP_h}{2J\Gamma}\pa{\frac{\tilde{\Phi}}{\sqrt{2}}\tilde{H}_+-W_0}$ &
	$\SL(2,\R)$ highest-weight $\U(1)$-invariant \\
	& $h\in\C$, $m=0$ \\
\hline
	& Type E \\
	$\tilde{A}_{h,m}^E(P_{h,m})=\displaystyle
	\frac{\tilde{\Phi}^h\tilde{\Psi}^mP_{h,m}}{2J\Gamma}\pa{\frac{\tilde{\Phi}}{\sqrt{2}}\tilde{H}_+-W_0}$ &
	$\SL(2,\R)$ highest-weight $\U(1)$-eigenstate \\
	& $h\in\cu{0,1}$, $m\in\Z$ \\
\hline
	& Type EM \\
	$\tilde{A}_{h,0}^{EM}(U_h)=\displaystyle
	\frac{\tilde{\Phi}^hU_h}{2J\Gamma}\br{\frac{\tilde{\Phi}}{\sqrt{2}}\pa{1+\frac{DhU_h^{-1/h}}{h-1}}\tilde{H}_+
		-\pa{1-\frac{1}{\Lambda^2}+\frac{DhU_h^{-1/h}}{h-1}}W_0+\frac{CU_h^{-1/h}}{\Lambda}\Theta}$ &
	$\SL(2,\R)$ highest-weight $\U(1)$-invariant \\
	& $h\in\C$, $m=0$ \\
\hline
	& Type EM \\
	$\tilde{A}_{h,m}^{EM}(U_{h,m})=\displaystyle
	\frac{\tilde{\Phi}^h\tilde{\Psi}^mU_{h,m}}{2J\Gamma}\cu{
		\frac{\tilde{\Phi}}{\sqrt{2}}\br{h(h-1)+m^2}\tilde{H}_+
		-\br{h(h-1)+m^2\pa{1-\frac{1}{\Lambda^2}}}W_0
		-im\frac{U_{h,m}'}{U_{h,m}}\Theta}$ &
	$\SL(2,\R)$ highest-weight $\U(1)$-eigenstate \\
	& $h\in\C$, $m\in\Z$ \\
\hline
\end{tabular}}
\caption{Complete classification of vector potentials $\tilde{A}$ that define solutions $\tilde{F}=\ed \tilde{A}$ to the force-free equations \eqref{eq:FFEtilde1}-\eqref{eq:FFEtilde3} that form representations of the $\SL(2,\R)\times\U(1)$ isometry group of NHEK. In every entry, it is understood that $\gamma=0$ in $\tilde{H}_+$.}
\label{table:Classification}
\end{table}

\subsection{An illustrative example}
\label{subsec:GlobalBasis}

Consider as another example the $\SL(2,\C)$ transformation obtained by setting
\begin{align}
	\alpha=\beta=\frac{i}{\sqrt{2}},
\end{align}
which sends the real basis $\cu{H_0,H_\pm}$ of $\SL(2,\R)$ into a complex basis
\begin{align}
	\cu{\tilde{H}_0,\tilde{H}_\pm}=\cu{L_0,\mp i\pa{\sqrt{2}e^\gamma}^{\mp1}L_\pm},
\end{align}
where $\cu{L_0,L_\pm}$ is precisely the global basis of $\SL(2,\R)$ used in our previous work \cite{Lupsasca:2014pfa},
\begin{align}
	L_0=\frac{i}{\sqrt{2}}\pa{\frac{1}{2}H_++H_-},\qquad
	L_{\pm}=\mp H_0+\frac{i}{\sqrt{2}}\pa{\frac{1}{2}H_+-H_-},\qquad
	Q_0=-iW_0.
\end{align}
It also sends $\Phi^h\Psi^m$, a scalar highest-weight representation of $\SL(2,\R)$ with respect to the $\cu{H_0,H_\pm}$ basis, into $\tilde{\Phi}^h\tilde{\Psi}^m$, a new scalar highest-weight representation of $\SL(2,\R)$ with respect to $\cu{L_0,L_\pm}$:
\begin{align}
	\tilde{\Phi}^h\tilde{\Psi}^m=\pa{4^he^{m\pi}}e^{im\phi+2m\arctan\pa{r+irt}}\br{\frac{r}{1-(t-i)^2r^2}}^h.
\end{align}
Knowing this, we can form the analogue of the axisymmetric Electric Type solution,
\begin{align}
	A_h^E(P_h)=\frac{\Phi^hP_h}{2J\Gamma}\pa{\frac{\Phi}{\sqrt{2}}H_+-W_0},
\end{align}
in this new basis of $\SL(2,\R)$ by following the procedure \eqref{eq:Procedure} -- the result is
\begin{align}
	\label{eq:OldSolution}
	\tilde{A}_h^E(P_h)
	=\frac{\tilde{\Phi}^hP_h}{2J\Gamma}\br{\frac{\tilde{\Phi}}{\sqrt{2}}\pa{e^\gamma\tilde{H}_+}-W_0}
	=-i\frac{\tilde{\Phi}^hP_h}{2J\Gamma}\pa{\frac{\tilde{\Phi}}{2}L_++Q_0}.
\end{align}
Recalling the transformation of the NHEK geometry to global coordinates,
\begin{align}
	r=\frac{\cos{\tau}-\cos{\psi}}{\sin{\psi}},\qquad
	t=\frac{\sin{\tau}}{\cos{\tau}-\cos{\psi}},\qquad
	\phi=\varphi+\ln\ab{\frac{\cos{\tau}-\sin{\tau}\cot{\psi}}{1+\sin{\tau}\csc{\psi}}}.
\end{align}
we see that in global coordinates $(\tau,\psi,\theta,\varphi)$, $\tilde{\Phi}$ becomes
\begin{align}
	\tilde{\Phi}(\tau,\psi)=2e^{-i\tau}\sin{\psi},
\end{align}
which agrees with the definition of $\tilde{\Phi}(\tau,\psi)$ in \cite{Lupsasca:2014pfa} up to a factor of 2 that cancels the extra factor in \eqref{eq:OldSolution}. Thus \eqref{eq:OldSolution} explicitly matches the solution described in \cite{Lupsasca:2014pfa}. Its physical properties, which we analyzed therein, are different from the ones of $A_h^E(P_h)$. In particular, recall from that paper that $\tilde{A}_h^E(P_h)$ had $F^2\neq0$, but with oscillating sign. Meanwhile $A_h^E(P_h)$ has $F^2<0$. Since $F^2$ is a geometric invariant of $\U(1)$ gauge fields, this proves the physical inequivalence of these solutions.

\subsection{General proof}

We wish to demonstrate that $\tilde{A}_{h,m}^M$, $\tilde{A}_{h,m}^E$ and $\tilde{A}_{h,m}^{EM}$ really solve the force-free equations \eqref{eq:FFE1}--\eqref{eq:FFE3} for all values of $\alpha,\beta\in\C$. This is in fact very simple, and we outline the steps of the proof.

First we recall that when $\alpha=\beta=0$, $\tilde{A}_{h,m}^M$, $\tilde{A}_{h,m}^E$ and $\tilde{A}_{h,m}^{EM}$ reduce to the Magnetic, Electric, and Electromagnetic Type solutions $A_{h,m}^M$, $A_{h,m}^E$ and $A_{h,m}^{EM}$, which were proved to solve the force-free equations in sections \ref{subsec:ProofTypeM}, \ref{subsec:ProofTypeE} and \ref{sec:TypeEM}, respectively.

Next, we consider $\alpha,\beta\in\R$. Turning on nonzero $\alpha$ and $\beta$ is equivalent to performing a finite $\SL(2,\R)$ transformation on these solutions. The resulting gauge potentials $\tilde{A}_{h,m}^M$, $\tilde{A}_{h,m}^E$ and $\tilde{A}_{h,m}^{EM}$ are therefore related by isometries to solutions of the force-free equations, from which it results that they are also (physically equivalent) solutions.

Finally, we observe that the force-free equations, when considered separately, are all linear. They are therefore insensitive to complexification, and it is in the sense we can call them ``holomorphic". As such, a complex finite $\SL(2,\R)$ transformation will still produce solutions, so $\tilde{A}_{h,m}^M$, $\tilde{A}_{h,m}^E$ and $\tilde{A}_{h,m}^{EM}$ must still solve the force-free equations even when $\alpha,\beta\in\C$.

To be more explicit, let's once again consider the case of the Magnetic Type solution $\tilde{A}_{h,m}^M$. Recall from \eqref{eq:TypeMstarF} and \eqref{eq:TypeMCurrent} that when $\alpha=\beta=0$, we find that the potential $\tilde{A}_{h,m}^M=A_{h,m}^M$ leads to
\begin{align}
	\J_{h,m}\propto H_+,\qquad\star F_{h,m}=H_+\wedge...
\end{align}
The force-free equations are then obviously satisfied because $\star F_{h,m}\wedge\J_{h,m}\propto H_+\wedge H_+=0$. Likewise, when $\alpha$ and $\beta$ are nonzero, we find that the more general potential $\tilde{A}_{h,m}^M$ leads to
\begin{align}
	\tilde{\J}_{h,m}\propto\tilde{H}_+,\qquad\star\tilde{F}_{h,m}=\tilde{H}_+\wedge...
\end{align}
The force-free equations are still obviously satisfied because $\star\tilde{F}_{h,m}\wedge\tilde{\J}_{h,m}\propto\tilde{H}_+\wedge\tilde{H}_+=0$.

\section{$\SL(2,\R)$-descendants and resummation of solutions}
\label{sec:Resummation}

In this section we examine the $\SL(2,\R)$-descendants of the solutions with highest weight (primaries). In the Poincar\'{e} basis, the descendants will no longer be annihilated by $H_+$, so they will acquire a time-dependence. Also, the electric or magnetic character of a primary is generally different than that of its solutions. For instance, the purely electric solutions have electromagnetic descendants.

\subsection{$\SL(2,\R)$-descendants as solutions}

First, we need to determine whether $\SL(2,\R)$-descendants are also solutions to the force-free equations \eqref{eq:FFE1}--\eqref{eq:FFE3}, as this is not automatically the case. Given a highest-weight solution $\tilde{A}_{h,m}$, its $k^\text{th}$ $\SL(2,\R)$-descendent is defined as
\begin{align}
	\tilde{A}_{h,k,m}=\L_{\tilde{H}_-}^k\tilde{A}_{h,m}.
\end{align}
In other words, descendants are obtained by acting with $\SL(2,\R)$ infinitesimally. Note that in contrast to the compact group $\SO(3)$, which has finitely many descendants that form the spherical harmonics $Y_{\ell,m}(\theta,\phi)$, the noncompact $\SL(2,\R)$ group generates an infinite tower of descendants, so $k\in\Z$.

Finite $\SL(2,\R)$ transformations (isometries) always transform solutions of an equation into other solutions. However, at the infinitesimal level, this is in general true only for linear equations. The force-free equations, being nonlinear, offer no guarantee of having this property and in fact they generally do not. They will nonetheless have it in certain circumstances, which we characterize in appendix \ref{appendix:Collinearity}. In particular, when descendants have current $\tilde{\J}_{h,k,m}=\L_{\tilde H_-}^k\tilde{\J}_{h,m}$ that is collinear to the current of the primary, $\tilde{\J}_{h,k,m}\parallel\tilde{\J}_{h,m}$, Theorem 1 from appendix \ref{appendix:Collinearity} guarantees that they still form solutions. Knowing this, we now examine the solutions of Type E, M and EM from Table \ref{table:Classification}.
\begin{itemize}
\item
Type M: these solutions have current $\tilde{\J}^M_{h,m}\propto\tilde{H}_+$. Since $\L_{\tilde{H}_-}\tilde{H}_+=\br{\tilde{H}_-,\tilde{H}_+}=-2\tilde{H}_0\not\propto\tilde{H}_+$, we see that the current of the Type M descendants is not collinear with that of the primary: $\tilde{\J}^M_{h,k,m}\not\,\parallel\tilde{\J}^M_{h,m}$. Therefore the Type M descendants are not solutions.
\item
Type E: these solutions have current $\tilde{\J}^E_{h,m}\propto W_0$. Since $\L_{\tilde{H}_-}W_0=\br{\tilde{H}_-,W_0}=0$, we see that the current of the Type E descendants is collinear with that of the primary: $\tilde{\J}^E_{h,k,m}\parallel\tilde{\J}^E_{h,m}$. Therefore the Type E descendants do form solutions.
\item
Type EM: these solutions have current $\tilde{\J}^{EM}_{h,m}$ proportional to a combination of $\tilde{H}_+$, $\tilde{H}_0$ and $W_0$, which changes direction under the action of $\tilde{H}_-$ (unless $h=1$ or $m=0$, in which case we retrieve a Type M or Type E solution, respectively). Hence the current of the Type EM descendants is not collinear with that of the primary: $\tilde{\J}^{EM}_{h,k,m}\not\,\parallel\tilde{\J}^{EM}_{h,m}$. Therefore the Type EM descendants are not solutions.
\end{itemize}

\subsection{General resummation of solutions}

More generally, linear superpositions of force-free solutions with collinear currents also form solutions. This is demonstrated in Theorem 3 of appendix \ref{appendix:Collinearity} for sums over $\SL(2,\R)$-descendants, and in Theorem 2 for arbitrary solutions. Hence we can infer that the following are solutions:
\begin{itemize}
\item
Linear combinations of Type M primaries $\tilde{A}_{h,m}^M$, since these solutions all have collinear currents $\tilde{\J}^M_{h,m}\propto\tilde{H}_+$. The same is not true of Type M descendants.
\item
Linear combinations of Type E primaries and their descendants $\tilde{A}_{h,k,m}^E$, since these solutions all have collinear currents $\tilde{\J}^E_{h,k,m}\propto W_0$. If $m=0$ then $h$ and $k$ are arbitrary, but if $m\neq0$ then either $h=0$ or $h=1$ (and $k$ is still arbitrary).
\end{itemize}
The Type EM solutions (both primaries and descendants) all have non-collinear currents and therefore do not admit a principle of linear superposition.

\section{Real solutions}
\label{sec:RealSolutions}

The solutions we have classified so far have nice representation-theoretic properties, but are not necessarily physical -- for instance, many of them stem from complex vector potentials and are therefore physically inadmissible. In this section, we extract real (physical) solutions from the classification given in Table \ref{table:Classification} and attempt to construct more general solutions. In particular, several solution subspaces admit a linear superposition principle. It is therefore possible to combine solutions with different highest weights, resulting in general solutions with no definite transformation properties. The resulting form of these solutions is akin to an expansion in ``conformal harmonics".

\subsection{Stationary solutions}

The highest-weight condition \eqref{eq:HW1} in the Poincar\'{e} basis amounts to demanding stationarity. Therefore, in order to obtain stationary solutions, we must only consider highest-weight primaries in this basis, for their $\SL(2,\R)$-descendants will yield time-dependent solutions, as will the highest-weight towers obtained from other $\SL(2,\R)$ embeddings.

\subsubsection{Stationary axisymmetric solutions}

For axisymmetric solutions, we further restrict our attention to $\U(1)$-invariant solutions with $m=0$. By taking the real and imaginary parts of $A_{h,0}^M$ and superposing them with different weights $h$, we obtain a very general stationary and axisymmetric solution of Magnetic Type:
\begin{align}
	A^M=\frac{1}{2J\Gamma}\br{\frac{\Phi}{\sqrt{2}}H_+-\pa{1-\frac{1}{\Lambda^2}}W_0}
	\int\!\ed h\br{c(h)r^{-h}+d(h)r^{h-1}}S_h,
\end{align}
where $c(h)$, $d(h)$ are arbitrary real functions, and $S_h(\theta)=S_{1-h}(\theta)$ is defined by \eqref{eq:ThetaTypeM}:
\begin{align}
	S_h''-\frac{\Lambda'}{\Lambda}\pa{\frac{1+\Lambda^2}{1-\Lambda^2}}S_h'+h(h-1)S_h=0.
\end{align}
The solution $S_h(\theta)$ is given in appendix \ref{appendix:ThetaAnalysis}. Because of its symmetry under $h\to1-h$, keeping both terms in the integrand is redundant. Without loss of generality, one may set $d(h)=0$, leaving
\begin{align}
	A^M=\frac{1}{2J\Gamma}\br{\frac{\Phi}{\sqrt{2}}H_+-\pa{1-\frac{1}{\Lambda^2}}W_0}
	\int\!\ed h\,c(h)\Phi^hS_h
\end{align}
as the resummed form of the magnetic primaries. Likewise, we can perform a resummation of the electric primaries, resulting in a very general stationary and axisymmetric solution of Electric Type:
\begin{align}
	A^E=\frac{1}{2J\Gamma}\br{\frac{\Phi}{\sqrt{2}}H_+-W_0}
	\int\!\ed h\br{c(h)r^{-h}+d(h)r^{h-1}}P_h
	\cong\frac{1}{2J\Gamma}\br{\frac{\Phi}{\sqrt{2}}H_+-W_0}
	\int\!\ed h\,c(h)\Phi^hP_h,
\end{align}
where $c(h)$, $d(h)$ are arbitrary real functions, and $P_h(\theta)=P_{1-h}(\theta)$ is defined by \eqref{eq:ThetaTypeE}:
\begin{align}
	P_h''+\frac{\Lambda'}{\Lambda}P_h'+h(h-1)P_h=0.
\end{align}
The solution $P_h(\theta)$ is given in appendix \ref{appendix:ThetaAnalysis}. As for the electromagnetic primaries, they are not linearly compatible and do not admit a superposition principle. The most general real, stationary and axisymmetric solution of Electromagnetic Type is therefore given by \eqref{eq:TypeEM}:
\begin{align}
	A_h^{EM}=\frac{\Phi^hU_h}{2J\Gamma}\br{\frac{\Phi}{\sqrt{2}}\pa{1+\frac{DhU_h^{-1/h}}{h-1}}H_+
		-\pa{1-\frac{1}{\Lambda^2}+\frac{DhU_h^{-1/h}}{h-1}}W_0+\frac{CU_h^{-1/h}}{\Lambda}\Theta},
\end{align}
with $U_h$ defined by \eqref{eq:ThetaTypeEM}.

\subsubsection{Stationary non-axisymmetric solutions}

By taking the real and imaginary parts of $A_{h,m}^M$ and superposing them with different weights $(h,m)$, we obtain a very general stationary non-axisymmetric solution of Magnetic Type:
\begin{align}
	A^M&=\frac{1}{2J\Gamma}\int\!\ed h\ed m\,c(h)\Phi^h\cu{b_mS_{h,m}
		\br{\frac{\Phi}{\sqrt{2}}H_+-\pa{1-\frac{1}{\Lambda^2}}W_0}
		-b_m'S_{h,m}'\chi_{h,m}\pa{1-\frac{1}{\Lambda^2}}\Theta},
\end{align}
where $\chi_{h,m}(\theta)$ is defined as in \eqref{eq:Chi},
\begin{align}
	\chi_{h,m}=\br{h(h-1)+m^2\pa{1-\frac{1}{\Lambda^2}}}^{-1},
\end{align}
while $b_m(\phi)$ is given by
\begin{align}
	b_m=x(m)\Re\Psi^m+y(m)\Im\Psi^m=x(m)\cos{m\phi}+y(m)\sin{m\phi}.
\end{align}
Here, $c(h)$, $x(m)$ and $y(m)$ are arbitrary real functions, and $S_{h,m}(\theta)=S_{1-h,m}(\theta)$ is defined by \eqref{eq:ThetaTypeM}:
\begin{align}
	S_{h,m}''+\frac{\Lambda'}{\Lambda}\br{1-\frac{2h(h-1)\chi_{h,m}}{1-\Lambda^2}}S_{h,m}'
	+\frac{S_{h,m}}{\chi_{h,m}}=0,
\end{align}
The solution $S_{h,m}(\theta)$ is given in appendix \ref{appendix:ThetaAnalysis}. As before, the symmetry under $h\to1-h$ allowed us to eliminate the $d(h)r^{h-1}$ term. This is the resummed form of the magnetic primaries. Likewise, we can perform a resummation of the electric primaries with $h=1$ or $h=0$, resulting in
\begin{align}
	A_1^E=\frac{\Phi}{2J\Gamma}\pa{\frac{\Phi}{\sqrt{2}}H_+-W_0}
	\int\!\ed m\br{x(m)\cos{m\phi}+y(m)\sin{m\phi}}P_m,
\end{align}
and
\begin{align}
	A_0^E=\frac{1}{2J\Gamma}\pa{\frac{\Phi}{\sqrt{2}}H_+-W_0}
	\int\!\ed m\br{x(m)\cos{m\phi}+y(m)\sin{m\phi}}P_m,
\end{align}
respectively. In both cases, $P_m$ is given by \eqref{eq:ThetaTypeEcharged} as
\begin{align}
	P_m(\theta)=\int\!\ed\theta\,e^{m/\Lambda(\theta)}
	=C\cosh\pa{\frac{m}{2}\cos\theta+\log\tan^m\frac{\theta}{2}}
	+D\sinh\pa{\frac{m}{2}\cos\theta+\log\tan^m\frac{\theta}{2}}.
\end{align}
In the electromagnetic case with $m\neq0$, we were unable to obtain any real solutions at all because taking the real and imaginary part of $\J_{h,m}^{EM}$ changes its direction.

\subsection{Time-dependent solutions}

Time-dependent solutions may be obtained in two ways: either by using $\SL(2,\C)$ automorphisms, or by acting with $H_-$ on the highest-weight primaries to obtain $\SL(2,\R)$-descendants which are no longer annihilated by $H_+$ and are therefore time-dependent. This section explores the latter method -- the automorphisms are then obtained following the procedure \eqref{eq:Procedure} elaborated in section \ref{sec:Automorphisms}.

First, by taking the real and imaginary parts of the $\SL(2,\R)$-descendants of Magnetic Type $\L_{H_-}^kA_{h,m}^M$ and superposing them with different weights $(h,k)$, we obtain a very general non-stationary solution,
\begin{align}
	A_t^M&=\frac{f}{2J\Gamma}\Theta,
\end{align}
where
\begin{align}
	f\pa{t\pm\frac{1}{r},\theta}
\end{align}
is a completely arbitrary function. Note that this solution is no longer of purely Magnetic Type, since time-varying magnetic fields produce electric fields -- instead, $A_t^M$ describes an electromagnetic field configuration. Also, observe that this solution is still axisymmetric, since descendants with $m\neq0$ are not resummable. This is thus the most general resummed form of the magnetic primaries\footnote{Incidentally, this is the only null solution ($F^2=0$) we can obtain in this approach. Unsurprisingly, it is related to the family of null solutions from \cite{Lupsasca:2014pfa} by the same $\SL(2,\C)$ automorphism that we already discussed in section \ref{subsec:GlobalBasis}.}. Likewise, we can perform a resummation of the electric descendants, resulting in
\begin{align}
	A_t^E(h)=\frac{\Phi^hP_h}{2J\Gamma}\br{\Phi\pa{\frac{h+1}{\sqrt{2}}tH_+-H_0}-htW_0},
\end{align}
where $P_h(\theta)=P_{1-h}(\theta)$ is defined by \eqref{eq:ThetaTypeE}:
\begin{align}
	P_h''+\frac{\Lambda'}{\Lambda}P_h'+h(h-1)P_h=0.
\end{align}
The solution $P_h(\theta)$ is given in appendix \ref{appendix:ThetaAnalysis}. Since this equation is only invariant under $h\to1-h$, we may only superpose solutions with highest weights $h$ and $1-h$:
\begin{align}
	A_t^E=CA_t^E(h)+DA_t^E(1-h),
\end{align}
for some arbitrary real coefficients $C$ and $D$. As for the electromagnetic descendants, we obtain
\begin{align}
	A_t^{EM}=\frac{g}{2J\Gamma}\br{\frac{C}{\Lambda}\Theta+\frac{D}{\Lambda^2}W_0},
\end{align}
where $C$ and $D$ are arbitrary real coefficients, while
\begin{align}
	g\pa{t\pm\frac{1}{r}}
\end{align}
is a completely arbitrary function on the AdS$_2$ subfactor of NHEK. As a final note, recall from section \ref{sec:Resummation} that the $\SL(2,\R)$-descendants of the Type M and Type EM highest-weight solutions do not form solutions. We evaded this limitation and were nonetheless able to obtain solutions $A_t^M$ and $A_t^{EM}$ by modifying the $\theta$-dependence of the complex highest-weight solutions.

\subsection{Survey of Solutionland}
\label{subsec:Survey}

In the previous two sections, we extracted real solutions of the force-free equations from our classification in Table \ref{table:Classification} of the highest-weight solutions in the Poincar\'e basis $\cu{H_0,H_\pm}$. We summarize these results in Table \ref{table:RealSolutions}.

A similar analysis may be performed for the highest-weight solutions with respect to other bases of $\SL(2,\R)$, such as the global basis $\cu{L_0,L_\pm}$ used in \cite{Lupsasca:2014pfa} and reviewed in section \ref{subsec:GlobalBasis}. All such solutions will be non-stationary, including the primaries, but they will be of the same form as the Poincar\'{e} basis solutions displayed in Table \ref{table:RealSolutions}.

They can be obtained by following the procedure \eqref{eq:Procedure} detailed in section \ref{sec:Automorphisms}, and replacing the coordinates $(t,r)$ by their analogues $(t',r')$ given in \eqref{eq:tprime}--\eqref{eq:rprime} wherever they appear. As an example, the Poincar\'{e} basis solution
\begin{align}
	A_t^{EM}=\frac{1}{2J\Gamma}g\pa{t\pm\frac{1}{r}}
		\br{\frac{C}{\Lambda}\Theta+\frac{D}{\Lambda^2}W_0}
\end{align}
belongs to an $\SL(2,\R)/\U(1)$ family of solutions parameterized by $(\alpha,\beta)$,
\begin{align}
	\tilde{A}_t^{EM}=\frac{1}{2J\Gamma}g\pa{t'\pm\frac{1}{r'}}
		\br{\frac{C}{\Lambda}\Theta+\frac{D}{\Lambda^2}W_0},
\end{align}
where
\begin{align}
	g\pa{t'\pm\frac{1}{r'}}=g\pa{-\frac{2r}{\br{\sqrt{2}(1+\alpha\beta)+\beta t}\beta r\pm\beta^2}}
\end{align}
is in general complex for $\alpha,\beta\in\C$. However, by current collinearity, we may take the real or imaginary part of $g$ and still obtain a real physical solution. The same can be done with the null ($F^2=0$) solution
\begin{align}
	A_t^M&=\frac{f}{2J\Gamma}\Theta.
\end{align}
In particular, transforming it to the global basis $\cu{L_0,L_\pm}$ of $\SL(2,\R)$ using the $\SL(2,\C)$ transformation discussed in section \ref{subsec:GlobalBasis}, it reproduces the family of null solutions first found in \cite{Lupsasca:2014pfa} provided
\begin{align*}
	f\pa{t'\pm\frac{1}{r'},\theta}=\pa{t'\pm\frac{1}{r'}}\frac{\tilde P(\theta)\Lambda(\theta)}{\sqrt{2}},
\end{align*}
with $\tilde P(\theta)$ an arbitrary function.

\subsection{Near-Horizon Near-Extremal Kerr black hole (near-NHEK)}
\label{subsec:nearNHEK}

\begin{figure}[t!]
	\begin{center}
		\includegraphics[width=.45\textwidth]{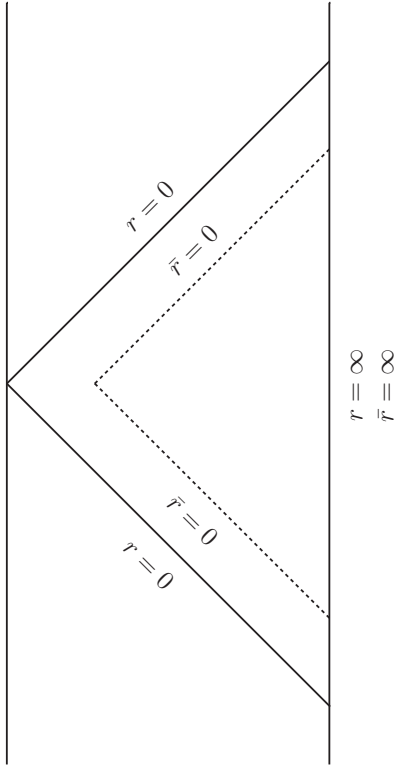}
	\end{center}
	\caption{Penrose diagram for the NHEK and near-NHEK geometries. The NHEK horizon (delineated by full lines) is at $r=0$, and lies beyond the near-NHEK horizon (delineated by dotted lines) at $\bar r=0$. The $r^{-h}$ singularities in our NHEK solutions are hidden behind the near-NHEK horizon.}
	\label{fig:PenroseDiagram}
\end{figure}

Finally, we would like to address the seemingly unphysical $r^{-h}$ singularities (for $h>0$) that pervade our force-free solutions in NHEK. By a conformal transformation, we can map the NHEK geometry to a spacetime describing the near-horizon region of a near-extremal Kerr black hole (near-NHEK). There are many such transformations, differing by the location of the near-NHEK horizon in relation to the NHEK horizon, as well as by the size of the deviation from extremality (or equivalently, the temperature $\kappa>0$ of the black hole). For our purposes, it is useful to consider the transformation
\begin{align}
	t=\frac{\sqrt{\bar r\pa{\bar r+2\kappa}}\sinh{\kappa\bar t}}
	{\bar r+\kappa+\sqrt{\bar r\pa{\bar r+2\kappa}}\cosh{\kappa\bar t}},\qquad
	r=\frac{\bar r+\kappa+\sqrt{\bar r\pa{\bar r+2\kappa}}\cosh{\kappa\bar t}}{\kappa},\\	\phi=\bar\phi-\frac{1}{2}\log\pa{\frac{\bar r}{\bar r+2\kappa}}
	+\log\pa{1-\frac{2\kappa}{\bar r+2\kappa+e^{\kappa\bar t}\sqrt{\bar r\pa{\bar r+2\kappa}}}},
\end{align}
which maps NHEK to the usual near-NHEK metric \cite{Maldacena:1998uz,Amsel:2009ev} in Poincar\'{e}-type coordinates $(\bar t,\bar r,\theta,\bar\phi)$,
\begin{align}
	ds^2=2J\Gamma\br{-\bar r\pa{\bar r+2\kappa}\ed t^2+\frac{\ed r^2}{\bar r\pa{\bar r+2\kappa}}
		+\ed\theta^2+\Lambda^2\br{\ed\phi+\pa{\bar r+\kappa}\ed t}^2}.
\end{align}
Unlike other transformations to near-NHEK (e.g. Eq.~(2.15) in \cite{Amsel:2009ev}), this particular one leads to a near-NHEK spacetime whose horizon $\bar r=0$ lies firmly outside the NHEK horizon, as illustrated by the Penrose diagram in Figure \ref{fig:PenroseDiagram}. As such, it maps our force-free solutions in NHEK to exact force-free solutions in near-NHEK, and in the process pushes the radial singularities in $\Phi^h(r)=r^{-h}$ behind the near-NHEK horizon. We defer the investigation of our force-free solutions in near-NHEK to future work.

\section*{Acknowledgements}
We are grateful to Andrew Strominger, Sam Gralla and Gim Seng Ng for useful conversations. This work was supported in part by NSF grant 1205550 and the Fundamental Laws Initiative at Harvard. M.J.R. is supported by the European Commission - Marie Curie grant PIOF-GA 2010-275082.

\appendix

\section{Analysis of $\theta$-dependence}
\label{appendix:ThetaAnalysis}

\subsection{Analysis of $P_h(\theta)$}

In order to determine the $\theta$-dependence of the highest-weight solutions of Type E, we need to determine the behavior of the function $P_h(\theta)$ and solve \eqref{eq:ThetaTypeE}, which may be rewritten as
\begin{align}
	\pd_\theta\pa{\Lambda\pd_\theta P_h}+h(h-1)\Lambda P_h=0.
\end{align}
This equation is manifestly invariant under $h\to1-h$, so $P_h(\theta)=P_{1-h}(\theta)$. By performing a suitable coordinate transformation to a new variable $z=\sin^2\theta$, we may put it in the form of a generalized Heun equation, namely
\begin{align}
	P_h''(z)+\pa{\frac{\gamma}{z}+\frac{\delta}{z-1}+\frac{\epsilon}{z-a}}P_h'(z)
	+\frac{\alpha\beta\pa{z-q}}{z(z-1)(z-a)}P_h(z)=0,
\end{align}
where $\alpha+\beta+1=\gamma+\delta+\epsilon$ and $q$ is an accessory parameter. In our case, these parameters are given by $\gamma=1$, $\delta=1/2$, $\epsilon=-1$, $\alpha=-h/2$, $\beta=(h-1)/2$ and $q=2$. The 4 regular singular points of this equation are located at $z=z_0$, with $z_0\in\cu{0,1,a=2,\infty}$. The corresponding roots $(t_1,t_2)$ of the indicial equation are $(0,1-\gamma)$, $(0,1-\delta)$, $(0,1-\epsilon)$ and $(\alpha,\beta)$, respectively.

We may solve \eqref{eq:ThetaTypeE} by expanding in a power series around each singularity $z_0$ (Frobenius' method). Since $z=\sin^2\theta$, the interval of interest to us, $\theta\in\br{0,\pi}$, gets mapped symmetrically to the interval $z\in\br{0,1}$; the poles $\theta=0,\pi$ are mapped to $z=0$, while the $\theta=\pi/2$ plane is mapped to $z=1$. Because there is no other singularity within this interval, the power series solutions obtained from the Frobenius method will converge everywhere on $z\in\br{0,1}$.

Expanding around $z_0=0$, we find that the two roots of the indicial equation are $(t_1,t_2)=(0,1-\gamma)$. Since $\gamma=1$, the roots are repeated: $t_1=t_2=0$. Hence only one of the two independent solutions to the equation is nonsingular. Though it has no closed form expression, it may be expanded as
\begin{align}
	P_h(\theta)=\sum_{n=0}^\infty a_n\sin^{2n}{\theta},
\end{align}
where
\begin{align}
	a_{n+1}=B_na_n+C_na_{n-1},
\end{align}
and
\begin{align}
	B_n=\frac{6n^2-h(h-1)}{4(n+1)^2},\qquad
	C_n=-\frac{(2n-h-2)(2n+h-3)}{8(n+1)^2}.
\end{align}
Per the above discussion, this power series converges everywhere on the domain of interest $\theta\in[0,\pi]$. Moreover, it renders manifest the reflection symmetry of $P_h$ about the $\theta=\pi/2$ plane. Finally, $B_n$ and $C_n$ are also invariant under $h\to1-h$, as expected.

\subsection{Analysis of $S_{h,m}(\theta)$}

In order to determine the $\theta$-dependence of the highest-weight solutions of Type M, we need to determine the behavior of the function $S_{h,m}(\theta)$ and solve \eqref{eq:ThetaTypeM}, which may be rewritten as
\begin{align}
	\pd_\theta\pa{\Lambda\pd_\theta S_{h,m}}-\frac{2h(h-1)\chi_{h,m}}{1-\Lambda^2}\Lambda'S_{h,m}'
	+\frac{\Lambda S_{h,m}}{\chi_{h,m}}=0.
\end{align}
The definition \eqref{eq:Chi} of $\chi_{h,m}(\theta)$ makes it manifestly invariant under $h\to1-h$ and $m\to-m$. Hence, this equation is manifestly invariant under these transformations, and $S_{h,m}(\theta)=S_{1-h,m}(\theta)$. By performing a coordinate transformation to a new variable $z=\sin^2\theta$, we may put it in the form
\begin{align}
	S_{h,m}''(z)+\frac{P(z)}{z}S_{h,m}'+\frac{Q(z)}{z^2}S_{h,m}=0,
\end{align}
where
\begin{align}
	P(z)&=\frac{m^2\pa{z^2-8z+4}^2\pa{z^2-6z+4}-4h(h-1)z^2\pa{3z^3-20x^2+52z-32}}
	{2(z-1)(z-2)\pa{z^2-8z+4}\br{m^2\pa{z^2-8z+4}-4h(h-1)z}},\\
	Q(z)&=\frac{m^2\pa{z^2-8z+4}-4h(h-1)z}{16(z-1)}.
\end{align}
This equation has 8 singularities, all of which are regular. They are located at $z=z_0$ with
\begin{align}
	z_0\in\cu{0,1,2,e_\pm=2\pa{2\pm\sqrt{3}},
	s_\pm=2\br{\Delta_{h,m}\pm\sqrt{\Delta_{h,m}^2-1}},\infty},
\end{align}
where $\Delta_{h,m}=2+h(h-1)/m^2$. We may solve \eqref{eq:ThetaTypeM} by expanding in a power series around each singularity $z_0$ (Frobenius' method). Since $z=\sin^2\theta$, the interval of interest to us, $\theta\in\br{0,\pi}$, gets mapped symmetrically to the interval $z\in\br{0,1}$; the poles $\theta=0,\pi$ are mapped to $z=0$, while the $\theta=\pi/2$ plane is mapped to $z=1$. The singularity at $z_0=e_-$ corresponds to the boundary of the ergosphere. Physically, we are therefore only interested in the interval $z\in\br{0,2\pa{2-\sqrt{3}}}$ outside the ergosphere, in which the only singularity we encounter is the one at $s_-$ (unless $h\in[0,1]$, in which case no singularity develops).

It therefore suffices for us to use the power series solutions around $z_0=0$ and $e_-$, which are respectively given by
\begin{align}
	S_{h,m}(z)=\sum_{n=0}^\infty c_nz^n,\qquad
	S_{h,m}(z)=\sum_{n=0}^\infty d_n(z-e_-)^n,
\end{align}
where the coefficients $c_k$ and $d_n$ are manifestly invariant under $h\to1-h$ and $m\to-m$ since
\begin{align}
	c_k&=-\frac{1}{k(k+m)}\cu{\frac{m^2}{16}c_{k-1}
		+\sum_{j=0}^{k-1}\br{(j+r)p(k-j)+\frac{m^2}{4}\pa{\Delta_{h,m}-\frac{5}{4}}}c_j},\\
	d_k&=-\frac{1}{k^2}\cu{\frac{m^2}{16}d_{k-1}
		+\sum_{j=0}^{k-1}\br{(j+r)q(k-j)+r(k-j)}d_j},
\end{align}
with
\begin{align}
	p(n)&=2^{-n}-\frac{1}{2}-\pa{\frac{e_+}{4}}^n-\pa{\frac{e_-}{4}}^n+\pa{\frac{s_+}{4}}^n+\pa{\frac{s_-}{4}}^n,\\
	q(n)&=\pa{-\frac{e_+}{8}}^nG_{n,h,m}-\pa{-e_-}^{-n}
	-\pa{\frac{1}{4\sqrt{3}}}^n\br{1+\frac{1}{2}\pa{2e_+}^n-\pa{\frac{e_+}{2}+1}^n},\\
	r(n)&=\frac{m^2}{4}\cu{\frac{\sqrt{3}}{2}\pa{\Delta_{h,m}-\frac{5}{4}}e_-\pa{1-e_-}^{-n}
		+\pa{-e_-}^{-n}\br{e_-\pa{\Delta_{h,m}-1}-(n-1)}},\\
	G_{n,h,m}&=\br{1+\frac{m^2}{h(h-1)}\pa{\sqrt{3}-\sqrt{\Delta_{h,m}^2-1}}}^n
		+\br{1+\frac{m^2}{h(h-1)}\pa{\sqrt{3}+\sqrt{\Delta_{h,m}^2-1}}}^n.
\end{align}
These two expansions will converge everywhere up to the singularity at $s_-$. They each have a free coefficient $c_0$ and $d_0$, and these must be matched to ensure smoothness across the singularity.

\section{Principle of superposition for collinear solutions of FFE}
\label{appendix:Collinearity}

\subsection{Infinitesimal symmetries of FFE}

The equations of FFE are nonlinear. As such, while their solutions are of course mapped to other solutions under finite symmetry transformations, this is not in general the case for infinitesimal symmetries. However, it is still sometimes possible, as explained by the following

\subsubsection*{Theorem 1:}
Suppose that $F$ is a solution to the system of equations \eqref{eq:HW1}--\eqref{eq:HW3}, and that $K$ is a Killing vector. Moreover suppose that $\L_K\J\propto\J$, or in other words that these currents are collinear: $\L_K\J\parallel\J$. (This condition is weaker than the assumption that $\J$ is an eigenstate of $K$, since the proportionality factor may be a function.) Then $\L_KF$ also solves equations \eqref{eq:FFE1}--\eqref{eq:FFE3}. Indeed, since $\br{\ed,\L_K}=\br{\ed^\dagger,\L_K}=0$,
\begin{align*}
	\ed\L_KF&=\L_K\ed F=0,
	\tag{\checkmark}\\
	\ed^\dagger\L_KF&=\L_K\ed^\dagger F=\L_K\J,
	\tag{\checkmark}\\
	\star\pa{\L_KF}\wedge\L_K\J&=\star F\wedge\J=0.
	\tag{\checkmark}
\end{align*}
The last line follows from the observation that\footnote{In this step we make use of the geometric identity $\iota_\eta\star\omega=\star\pa{\omega\wedge\eta}$, which holds of any 2-form $\omega$ and 1-form/vector $\eta$. Together with Hodge duality, it implies that in 4-dimensional Lorentzian spacetimes, $\iota_\eta\omega=-\star\pa{\star\omega\wedge\eta}$.}
\begin{align}
	-\star\br{\star\pa{\L_KF}\wedge\L_K\J}&=\iota_{\L_K\J}\L_KF
	\propto\iota_\J\L_KF=\br{\iota_\J,\L_K}F+\L_K\cancel{\iota_\J F}
	=\iota_{\br{\J,K}}F=-\iota_{\L_K\J}F\propto\iota_\J F\nonumber\\
	&=-\star\br{\star F\wedge\J}=0.
\end{align}
Note that every assumption is used in the proof; in particular, if $K$ is not a Killing vector then $\br{\ed^\dagger,\L_K}\neq0$ and the argument breaks down because $\ed^\dagger\L_KF\neq\L_K\J$. $\hfill\square$

\subsection{Principle of linear superposition}

Nonlinear equations such as those of FFE do not generally allow for a principle of linear superposition. However, as already noted in \cite{Lupsasca:2014pfa} and again in this paper, the equations of FFE sometimes admit linear combinations. This is explained by the following

\subsubsection*{Theorem 2:}
Suppose that $F_1$ and $F_2$ are solutions to the system of equations \eqref{eq:FFE1}--\eqref{eq:FFE3}, with associated currents $\J_1$ and $\J_2$, respectively. Moreover suppose that $\J_1\propto\J_2$, or in other words that these currents are collinear: $\J_1\parallel\J_2$. Then the linear combination $aF_1+bF_2$ also solves equations \eqref{eq:FFE1}--\eqref{eq:FFE3} for any constants $a$ and $b$. Indeed, using the assumption that $\J_1\propto\J_2$, so that $\iota_{\J_1}F_2\propto\iota_{\J_2}F_2=0$ and $\iota_{\J_2}F_1\propto\iota_{\J_1}F_1=0$, we see that
\begin{align}
	-\star\br{\star\pa{aF_1+bF_2}\wedge\pa{a\J_1+b\J_2}}&=\iota_{a\J_1+b\J_2}\pa{aF_1+bF_2}\\
	&=a^2\cancel{\iota_{\J_1}F_1}+b^2\cancel{\iota_{\J_2}F_2}+ab\pa{\iota_{\J_1}F_2+\iota_{\J_2}F_1}
	=0\nonumber.
\end{align}
Hence, we find that
\begin{align*}
	\ed\pa{aF_1+bF_2}&=a\ed F_1+b\ed F_2=0,
	\tag{\checkmark}\\
	\ed^\dagger\pa{aF_1+bF_2}&=a\ed^\dagger F_1+b\ed^\dagger F_2=a\J_1+b\J_2,
	\tag{\checkmark}\\
	\star\pa{aF_1+bF_2}\wedge\pa{a\J_1+b\J_2}&=\iota_{a\J_1+b\J_2}\pa{aF_1+bF_2}=0,
	\tag{\checkmark}
\end{align*}
as claimed. $\hfill\square$\vspace{5mm}

Theorem 2 explains why the primaries of different towers, which are not related by an infinitesimal symmetry but rather differ by their highest weight, can also be superposed as long as their currents are collinear. But we also need to explain why towers of descendants can be superposed. This is explained by the following

\subsubsection*{Theorem 3:}
(This is a generalization of Theorem 1.)

Suppose that $F$ is a solution to the system of equations \eqref{eq:FFE1}--\eqref{eq:FFE3}, and that $K$ is a Killing vector. Moreover suppose that $\L_K\J\propto\J$, or in other words that these currents are collinear: $\L_K\J\parallel\J$. Then for any polynomial of arbitrary degree $d\in\N$,
\begin{align}
	\P(x)=\sum_{n=0}^da_nx^n,	
\end{align}
the 2-form $\P\pa{\L_K}F$ also solves \eqref{eq:FFE1}--\eqref{eq:FFE3}. The proof follows from Theorems 1 and 2. $\hfill\square$\vspace{5mm}

The upshot of this analysis is that force-free solutions with collinear currents can be linearly superposed. Other interesting implications of current collinearity in FFE have been explored in \cite{Smolic:2014swa}.


\bibliography{FFENHEK}
\bibliographystyle{utphys}

\end{document}